\documentclass[twocolumn]{aastex61}
\usepackage{longtable}
\usepackage{graphicx}
\usepackage{amsmath,amssymb}
\usepackage{booktabs}
\usepackage{color}
\usepackage{units}
\usepackage{epstopdf}
\usepackage[caption=false]{subfig}
\usepackage{float}

\newcommand{\mej}{\ensuremath{M_{\text{ej}}}}

\newcommand{\beq}{\begin{equation}}
\newcommand{\eeq}{\end{equation}}
\newcommand{\bdm}{\begin{displaymath}}
\newcommand{\edm}{\end{displaymath}}

\definecolor{Gray}{gray}{0.9}
\definecolor{orange}{rgb}{0.9,0.5,0}
\definecolor{dkGreen}{rgb}{0.09, 0.45, 0.27}
\definecolor{purple}{rgb}{0.44, 0.0, 1.0}

\begin{document}
\title{Estimating the Contribution of Dynamical Ejecta in the Kilonova Associated with GW170817}
\author{B.~P.~Abbott}
\affiliation{LIGO, California Institute of Technology, Pasadena, CA 91125, USA}
\author{R.~Abbott}
\affiliation{LIGO, California Institute of Technology, Pasadena, CA 91125, USA}
\author{T.~D.~Abbott}
\affiliation{Louisiana State University, Baton Rouge, LA 70803, USA}
\author{F.~Acernese}
\affiliation{Universit\`a di Salerno, Fisciano, I-84084 Salerno, Italy}
\affiliation{INFN, Sezione di Napoli, Complesso Universitario di Monte S.Angelo, I-80126 Napoli, Italy}
\author{K.~Ackley}
\affiliation{University of Florida, Gainesville, FL 32611, USA}
\affiliation{OzGrav, School of Physics \& Astronomy, Monash University, Clayton 3800, Victoria, Australia}
\author{C.~Adams}
\affiliation{LIGO Livingston Observatory, Livingston, LA 70754, USA}
\author{T.~Adams}
\affiliation{Laboratoire d'Annecy-le-Vieux de Physique des Particules (LAPP), Universit\'e Savoie Mont Blanc, CNRS/IN2P3, F-74941 Annecy, France}
\author{P.~Addesso}
\affiliation{University of Sannio at Benevento, I-82100 Benevento, Italy and INFN, Sezione di Napoli, I-80100 Napoli, Italy}
\author{R.~X.~Adhikari}
\affiliation{LIGO, California Institute of Technology, Pasadena, CA 91125, USA}
\author{V.~B.~Adya}
\affiliation{Max Planck Institute for Gravitational Physics (Albert Einstein Institute), D-30167 Hannover, Germany}
% ...
\collaboration{LIGO Scientific Collaboration and Virgo Collaboration}




\def\BNSrateMin{\ensuremath{320}}
\def\BNSrateMax{\ensuremath{4740}}
\def\BNSrateMedian{\ensuremath{1540}}
\def\BNSrateCOMPACT{\ensuremath{\BNSrateMedian^{+3200}_{-1220}}}
\def\MinLLX{\ensuremath{10^{-7.2}}}
\def\MaxULX{\ensuremath{10^{-5.7}}}
\def\MedianX{\ensuremath{\sim 10^{2.5} {\rm  M}_\odot {\rm Mpc}^{-3}  }}
\def\MinLLrho{\ensuremath{10^{1.7}}}
\def\MaxULrho{\ensuremath{10^{3.2}}}
\def\frpmin{\ensuremath{10\%}}

\begin{abstract}

The source of the gravitational-wave signal GW170817, very likely a binary neutron star merger, was also observed electromagnetically, providing the first multi-messenger observations of this type. The two week long electromagnetic counterpart had a signature indicative of an r-process-induced optical transient known as a kilonova.
This Letter examines how the mass of the dynamical ejecta can be estimated without a direct electromagnetic observation of the kilonova, using gravitational-wave measurements and a phenomenological model calibrated to numerical simulations of mergers with dynamical ejecta.  
Specifically, we apply the model to the binary masses inferred from the gravitational-wave measurements, and use the resulting mass of the dynamical ejecta to estimate its contribution (without the effects of wind ejecta) to the corresponding kilonova light curves from various models.
The distributions of dynamical ejecta mass range between $\mej = 10^{-3} - 10^{-2}\mathrm{M}_\odot$ for various equations of state, assuming the neutron stars are rotating slowly. In addition, we use our estimates of the dynamical ejecta mass and the neutron star merger rates inferred from GW170817 to constrain the contribution of events like this to the r-process element abundance in the Galaxy when ejecta mass from post-merger winds is neglected. We find that if $\gtrsim10\%$ of the matter dynamically ejected from BNS mergers is converted to r-process elements, GW170817-like BNS mergers could fully account for the amount of r-process material observed in the Milky Way. 

\end{abstract}

\section{Introduction}
\label{sec:intro}
On August 17, 2017, 12:41:04 UTC, the LIGO -- Virgo gravitational-wave (GW) observatory network, composed of LIGO Hanford Observatory, LIGO Livingston Observatory, and Virgo, recorded GWs consistent with a binary neutron star (BNS) inspiral and merger \citep{LVC_only_paper}. This signal was subsequently named GW170817.

In addition to the GW signature, the merger of a BNS system is expected to have multiple electromagnetic signatures over different time scales \citep{Nakar2007,MeBe2012}.
The LIGO-Virgo sky localization of GW170817~\citep{LVC_only_paper} spurred an intensive multi-messenger campaign covering the whole electromagnetic spectrum to search for counterparts (see \citealt{MMAPaper} for an extended list). Within hours, broadband observations --- backed by archival data investigation --- revealed an optical transient~\citep*{SWOPE,DECam,DLT40,LASCUMBRES,TanvEA2017,PiDa2017,MASTER}, a type of transient called a kilonova~\citep{LiPa1998,Me2017} originating from neutron-rich matter unbound from the system \citep[e.g.,][]{SWIFT,McCully2017,Smartt2017,TrPi2017}.

Broadly, two types of ejecta are expected to contribute to kilonovae: dynamical ejecta produced at the time of merger~\citep{RosswogLiebendorfer99,MeMa2010,RoKa2011,Ro2013,Baka2013,TaHo2013,HoKi2013,BaGo2013,SeKi2016,Radice:2016dwd,Dietrich:2016hky,DiUj2017,BoMa2017}, and post-merger winds produced by the remnant system, for example from an accretion disk around a black hole or massive neutron star ~\citep{DeOt2009,PeRo2014,Martin:2015hxa,KiSe2015,FeKa2015,KaFe2015,FoOC2016,ShKi2017,SiMe2017,CiKa2017,FuSe2017}.

Both electromagnetic and GW measurements rely on models to connect the underlying properties and composition of the ejecta to their respective observations.  The process of interpreting ejecta based on electromagnetic observations  is described in \cite{Alexander17,LASCUMBRES,Chornock17,CoWi2017,Cowperthwaite17,DiMa2017,MAGELLAN,SWIFT,KaEA2017,McCully2017,Nicholl17,PiDa2017,Smartt2017,TaUt2017,TrPi2017,MMAPaper}.  We use phenomenological calculations that estimate the dynamical ejecta mass from the pre-coalescence binary properties, which GW observations can constrain.  This mass is a critical ingredient needed to predict contribution of dynamical ejecta to the EM light curve associated with this kilonova transient.  Going forward, this procedure would also assist in the interpretation of future followup observations where a dim counterpart was detected, or none at all.

This Letter shows how dynamical ejecta masses obtained from GW parameter estimates of GW170817 via phenomenological fits to numerical models for the mass and velocity of dynamically ejected matter in BNS systems~\citep{DiUj2017} (hereafter DU17) can predict kilonova light curves.
Similar numerical work has produced fitting formulae in the case of neutron-star black-hole (NSBH) binaries~\citep{KaKy2016}. While the GW detection of GW170817 cannot rule out the presence of a black hole companion, the BNS interpretation is favored~\citep{LVC_only_paper}. Consequently, we do not include the NSBH scenario in this work, and only employ the fitting formulas for ejecta mass and velocity from BNS simulations (DU17). The GW170817 analysis extracted the BNS source parameters using Bayesian inference~\citep{LVC_only_paper}, and those results are used here to estimate the mass of the dynamical ejecta. This approach accounts for the dependence of the amount of ejected matter on the size and stiffness~\citep{KaKy2016} of the components of the binary, characterized by the equation of state (EOS) and its influence on the mass-radius relationship~\citep{LaPr2001,OzFr2016}.

Bayesian inference with a GW signal model applied to the strain data provides a posterior distribution of component masses ($m_i$) and dimensionless spins ($\chi_i\equiv c|\mathbf{S}_i|/(Gm_i^2)$, where $\mathbf{S}$ is the angular momentum of the NS) consistent with the observations~\citep{PhysRevD.91.042003}. Assuming neutron stars spins are small ($\chi\leq 0.05$, hereafter ``low spin''), we obtain distributions of ejecta between $10^{-3}$ and $10^{-2}~\mathrm{M}_{\odot}$. Allowing for larger neutron stars spins ($\chi\leq 0.89$, hereafter ``high spin'') pushes some ejecta values higher, of the order of~$10^{-1} \mathrm{M}_{\odot}$ at its highest. In this Letter, we focus on dynamical sources, so it is important to recall that this analysis may not account for a significant fraction of the ejecta mass; winds could produce comparable or even more ejecta than from dynamical sources. Using the GW-derived dynamical ejecta estimates, the derived light curves vary significantly between the adopted models, in both color evolution and time and magnitude of peak emission; in extreme cases, they can reach beyond 15th magnitude in optical bands. 

Like supernovae~\citep{TeSu2001}, neutron star mergers are believed to contribute to the abundance of heavy elements \citep{LaSc1974} through r-process nucleosynthesis~\citep{BuBu1957}. If so, the frequency of kilonova events should then be intimately tied to the overall abundance of r-process generated material~\citep{LaSc1974}. 
Using our GW estimates of dynamical ejecta masses and the merger rates inferred from the BNS discovery (\BNSrateCOMPACT~Gpc${}^{-3}$yr${}^{-1}$)~\citep{LVC_only_paper}, we estimate a present-day r-process density of $\MinLLrho{}-\MaxULrho{}$ M$_\odot$Mpc$^{-3}$ contributed by BNS mergers.
Under the assumption that all BNS mergers produce the same amount of dynamical ejecta that we infer for GW170817, this estimate is consistent with the Galactic values and suggests the associated nucleosynthesis is one of the primary contributors to r-process abundances.

\section{Predicted Dynamical Ejecta Mass}
\label{sec:kN_ejecta}
The amount of ejecta from binary mergers in general depends on the masses and EOS of the two components, their rotation, and, most importantly for post-merger winds, the neutrino/radiation hydrodynamics and the magnetic fields, e.g.~\cite{HoKi2013,Martin:2015hxa,SeKi2016,Radice:2016dwd,Dietrich:2016hky,SiMe2017}. Based on detailed numerical studies of merging, irrotational binaries, the phenomenological fits devised by DU17 relate the dynamical ejecta mass \mej\ to the gravitational mass of the component stars ($m$), their baryonic mass ($m_b$), and their radii $R$ (or equivalently compactnesses $C = Gm/Rc^2$).  Contributions due to winds were not included in the simulations used by DU17, and thus are not part of the fits for \mej, even though they may lead to comparable ejecta masses.

Because the EOS in neutron stars is poorly constrained, two approaches are taken to describe the bulk properties of the binary components. In the first approach, we assume an EOS and infer $m_b$ and $C$ from the binary's measured gravitational masses using a zero-temperature non-rotating model (computed using the Oppenheimer-Volkoff equations, \citealt{OpVo1939}). Different EOS will predict different radii and baryonic masses for the same gravitational masses and, as such, will affect the amount of ejecta and the predicted light curve of the kilonova. The EOS of cold, dense, degenerate matter is poorly constrained (see \citealt{OeHe2017} for a recent review), so we evaluate a representative selection of the EOS considered in \cite{OzFr2016}. The tidal deformabilities allowed by GW170817~\citep{LVC_only_paper} do disfavor stiffer EOS; however, many remain compatible with our measurements. Due to observational constraints, we restrict ourselves to EOS that have a maximum mass above 1.97 $\text{M}_\odot$ \citep{DePe2010,AnFr2013}. Specifically, we consider EOS calculations from~\cite{Gl1985}, GNH3; \cite{MuPr1987}, MPA1; \cite{WiFi1988}, WFF1-2; \cite{EnBa1996}, ENG; \cite{MuSe1996}, MS1, MS1b; \cite{AkPa1998}, APR3-4; \cite{DoHa2001}, SLy; and \cite{LaNa2006}, H4.

In the second case, we take an approach that does not assume a specific EOS to compare against our EOS specific results. The internal structure of the neutron stars in a binary is encoded in the gravitational waveform through the (dimensionless) tidal deformabilities (denoted $\Lambda$) of the neutron stars \citep{FlHi2008,DaNa2012,DeLi2013,WaCr2014}. One can infer $m_b$ and $C$ from the binary's measured gravitational masses and tidal deformabilities by applying fits from \cite{CoDi2017} and \cite{YaYu2017}, which give $m_b(m,C)$ and $C(\Lambda)$, respectively. While some error is incurred using these additional fits, it is small compared to the estimated uncertainty of the fits for the dynamical ejecta properties and the intrinsic uncertainty in current numerical relativity simulations. Specifically, for the EOS considered by \cite{YaYu2017}, the error in the tidal deformability-compactness relation is $< 10\%$ for nuclear EOS, while for the baryonic mass fit, the maximum error found by \cite{CoDi2017} is $< 3\%$. When applying these fits, we also exclude cases with component masses above 3 $\text{M}_\odot$, a standard upper bound on neutron star masses \citep{KaBa1996}, and restrict the compactness to be below the Buchdahl bound \citep{Buc1959} of $4/9 \simeq 0.44$, which similarly only affects a few cases.

\subsection{Sources of Uncertainties in Ejecta Mass Estimation}
\label{sec:fit_uncertainty}

Many caveats must be considered when assessing the uncertainty in estimates of \mej. The amount of ejecta from mergers also depends on various microphysics, such as the particular treatment of thermal effects, neutrino transport, and magnetic fields~\citep{DeOt2008,PeRo2014,BaGo2013,SeKi2016,Radice:2016dwd,BoMa2017,CiKa2017}, which lead to uncertainties about the ejecta's structure, angular distribution, and composition \citep{KaBa2013,TaHo2013,BaKa2016}. These parameters are not included in the \mej\ fits in DU17. Additionally, the DU17 fits ignore the effects of spin on dynamical ejecta, which can change the amount of ejecta \citep{Kastaun:2014fna,DiBe2017,Kastaun:2016elu}. In particular, aligned spin can increase torque in the tidal tail and lead to more ejecta, which is most notable for unequal mass configurations. To understand the effect of spin on dynamical ejecta, additional better resolved simulations are needed.

Systematic uncertainties are also of concern. The accuracy of the \mej\ fit from DU17 relies on the underlying numerical relativity simulations. Simulation choices for input physics (nuclear EOS and microphysics), inclusion of different neutrino transport models, and chosen grid resolution can all result in large systematics. For example, comparison of numerical relativity predictions of \mej\ differ by a factor of ${\sim} 4$ \citep{SeKi2016,LeLi2016,BoMa2017}. Further, the error on ejecta masses from numerical simulations likely has an absolute component, leading to increasing relative errors for low ejecta masses --- for additional discussion see~\cite{EnCi2016,CiKa2017}. The error at low masses is not symmetric since \mej\ cannot be negative, potentially biasing the phenomenological fits of DU17 to overestimate the ejecta mass. Additionally, there are also systematic uncertainties introduced by the specific form of the fit, where all EOS effects are contained in the values of $m_b$ and $C$ for a given $m$. Finally, as discussed in \cite{LVC_only_paper} and Sec.~\ref{sec:kN_ej_results}, the waveform model used to infer the masses and tidal deformabilities from the gravitational wave signal introduces its own systematic uncertainties, though these are estimated to be smaller than those of the DU17 \mej\ fit.

All these considerations will contribute to the  uncertainty in the \mej\ fit from DU17; the error is a mixture of systematic errors that need investigation with dedicated future studies and numerical simulations. To model some part of this error, we will treat the average relative error of the fit quoted in DU17 (72\%) as a statistical error for any results used here and defer a more robust error analysis to future work. We include an estimate of the error of the \mej\ fit from DU17 by replacing each ejecta mass sample with a random value consistent with a Gaussian distribution in $\log_{10}\mej$ centered on the value and with standard deviation of $\log_{10} 1.72$, as motivated in Section \ref{sec:fit_uncertainty}. This method excludes zero ejecta masses and errors for small ejecta masses $\lesssim 10^{-3}\,M_\odot$ are not well modeled.
The ejecta mass fit is based on simulations with non-zero ejecta mass. The full parameter space likely also contains cases with little or no ejecta mass, for example, systems exhibiting prompt black hole formation. Since we reported in~\cite{GRBpaper} that prompt collapse can only be excluded for extreme EOS such as MS1, and the fit at values below $3\times 10^{-3} \, M_\odot$ strongly overestimates the ejecta mass compared to the NR data points, the fit cannot reliably exclude zero ejecta mass below this value.
Figure \ref{fig:ejecta_props} shows that, in the low-spin cases, the number of samples less than $\mej < 3\times 10^{-3} \mathrm{M}_{\odot}$ is typically $\sim10-15$\% of the cumulative total for most. In extreme cases, this fraction is up to ~50\%, but also arises from EOS which have been disfavored in~\cite{LVC_only_paper}. In the high-spin cases, this number is typically smaller, around 5-10\%, but can reach up to 25\% in the extreme cases. We also discard the few samples when the fit predicts a negative value.

\subsection{Ejecta Mass Predictions}
\label{sec:kN_ej_results}

We evaluate the \mej\ fit using the binary parameters derived from the gravitational-wave analysis \citep{LVC_only_paper}. These parameters include the gravitational masses, tidal deformabilities, and spins of the component stars, though the spins are not used in evaluating the fit. Bayesian inference provides a distribution of these parameter values as a set of independent samples drawn from the posterior~\citep{PhysRevD.91.042003,GW150914PE}. As a quantity that derives from these binary parameters, \mej\ then is also represented as a statistical sample.

While the estimation of \mej\ does not include the component spins as an input, they are an important degree of freedom in the waveform models used in the GW analysis. We consider two sets of GW parameter samples, defined by the choices for the prior on the spin magnitude. The two spin priors considered here are $\chi\leq 0.89$ (our ``high spin'' case with the upper limit dictated by the waveform model used), and $\chi\leq 0.05$ (our ``low spin'' case, slightly above the largest inferred spin at merger of a neutron star in a binary neutron star system that will merge within a Hubble time~\citep{Bu2003}). While the waveform models used only include the effects of the spin components along the orbital angular momentum, the spin priors assume isotropic spin directions. The very highest spins allowed in the high spin posterior set exceed the mass-shedding limit ($\chi\sim 0.7$ for the EOS considered in \citealt{LoLi2011}), but the small density of posterior samples in this region lies outside the 90\% credible intervals. More importantly, the high-spin posterior on the primary mass contains samples with masses well above the maximum mass allowed for a static NS for any of the EOS we consider; we simply exclude from consideration any samples with such unsupported masses for each EOS.

There are also systematic errors introduced by the waveform model used. As discussed in \cite{LVC_only_paper}, analysis with a different waveform model changes the 90\% credible bounds on the masses by $\sim 15\%$ in the high-spin case (with no changes in the low-spin case), and the bounds on the tidal deformabilities by $\sim 20$--$30\%$ in both low- and high-spin cases. Since these differences are below the systematic errors of the DU17 fit, we do not attempt to account for them here. The true systematic errors from waveform models may be significantly larger than those estimated in this comparison; making such assessments is the subject of future work. 

\begin{figure}[ht]\centering
\includegraphics[width=0.95\linewidth]{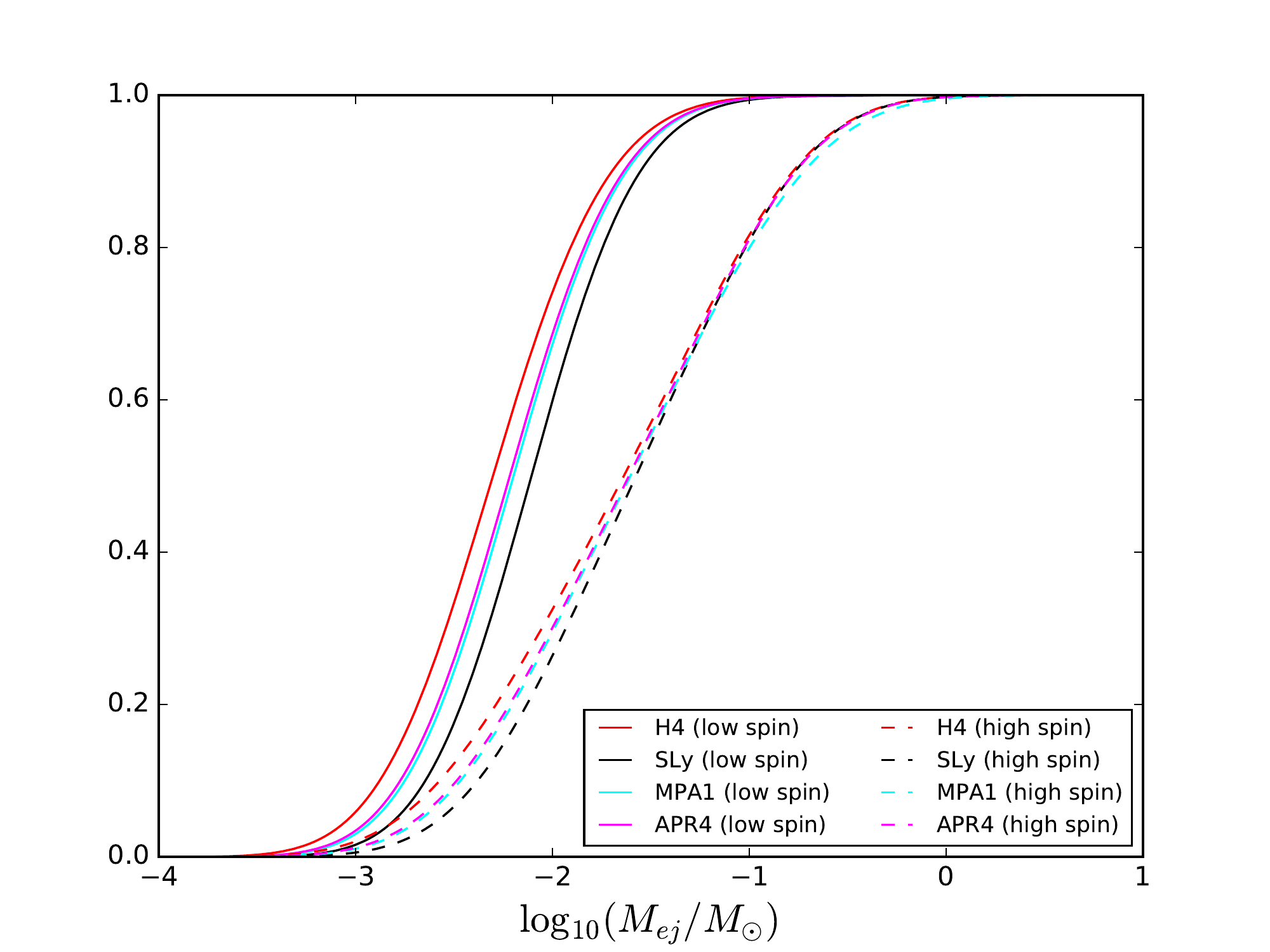}
\caption{The figure above displays the cumulative distribution function of the dynamical ejecta mass predicted for a representative selection of the EOS in the study. The low spin case are traced in solid colors and the high spin case are dashed curves.}
\label{fig:ejecta_props}
\end{figure}

Figure \ref{fig:ejecta_props} reports cumulative probability distributions for the dynamical ejecta for a selection of the EOS tested.
While all cases predict ejecta concentrated between $10^{-3}$ -- $10^{-2}\textrm{M}_{\odot}$, the high spin results allow for larger median ejecta values in general --- maximum values can exceed a tenth of a solar mass. Since the DU17 fits for \mej\ neglect spin, the differences in ejecta for the cases shown in Figure \ref{fig:ejecta_props} are driven by the imprint of the spin choices inherent in the GW analysis that was input into this analysis. 

\begin{figure*}[ht]\centering
\includegraphics[width=0.45\linewidth]{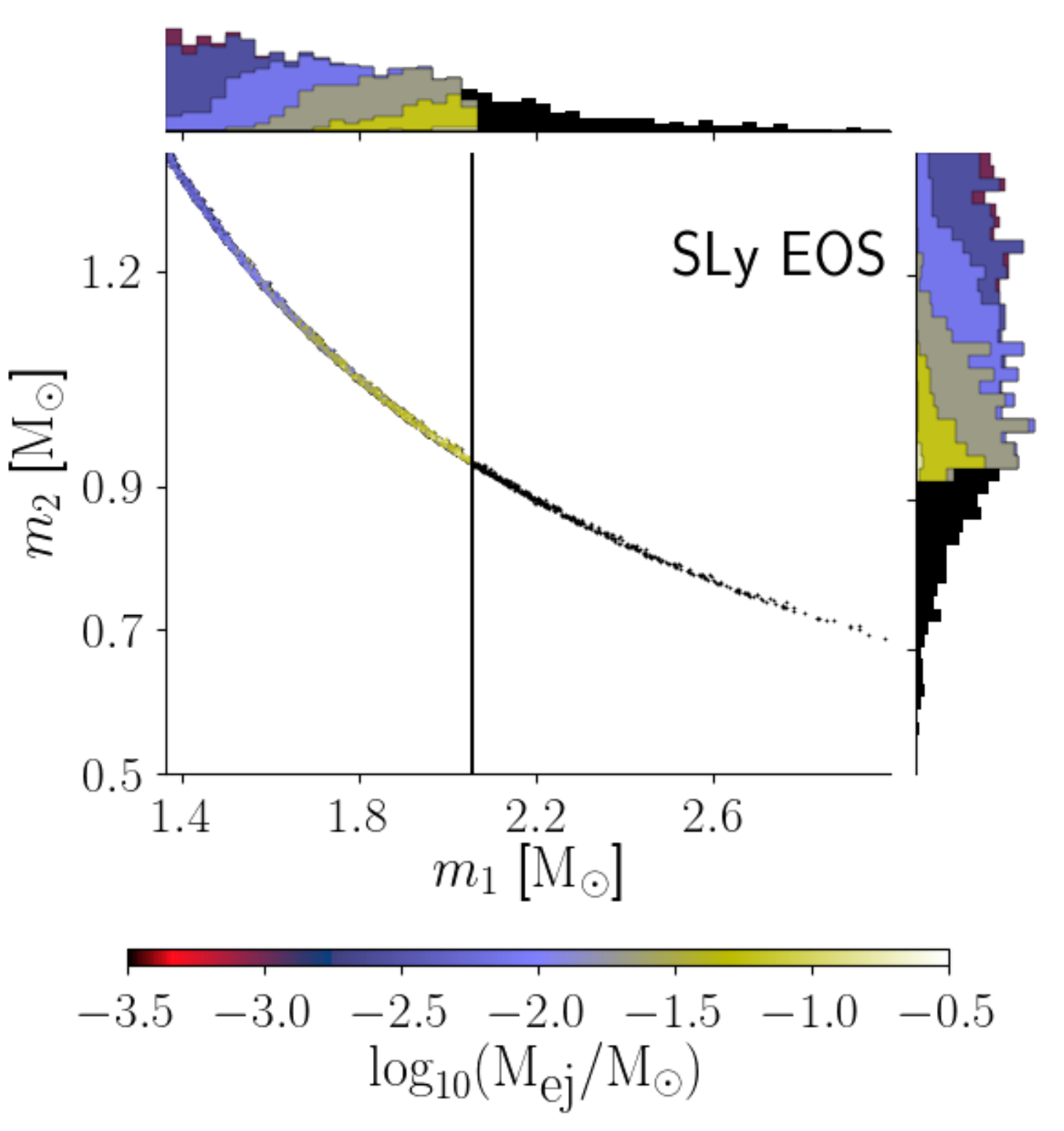}
\includegraphics[width=0.45\linewidth]{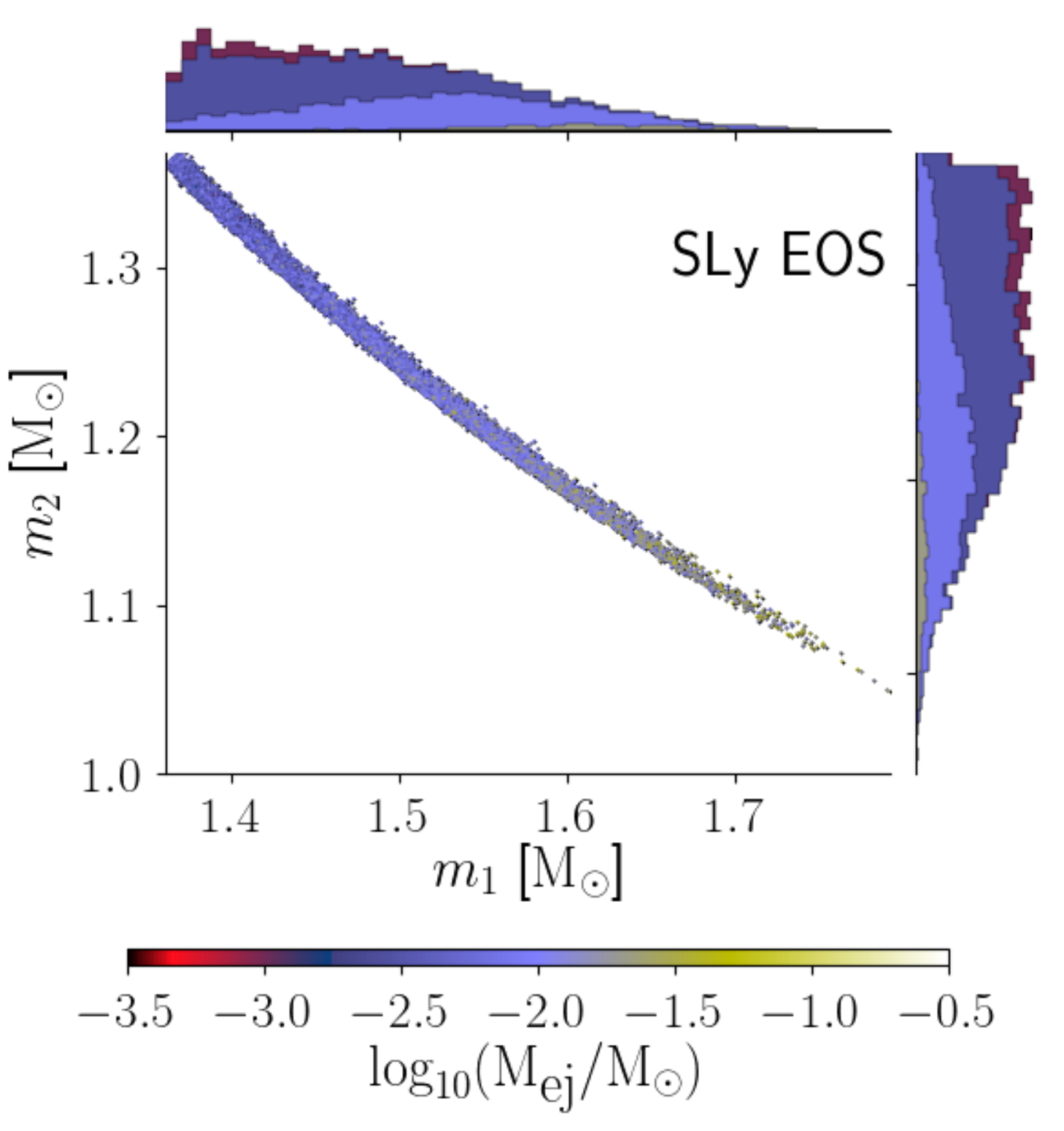}
\caption{The left (high-spin prior) and right (low-spin prior) panels above show the distribution of the primary ($m_1$) and secondary ($m_2$) masses from GW measurements. The color of each point indicates the predicted dynamical ejecta mass for each sample that the SLy EOS allows. In the left-hand plot, black markers correspond to $m_1$ values that are disallowed by the maximum mass of the EOS (marked by a vertical line). The underlying black histograms to the top and right of each plot are the one-dimensional marginalized histograms of the masses. The stacked histograms on top of them in various colors show the binary masses that create ejecta masses above logarithmically spaced thresholds of $1\times10^{-3}$, $3\times 10^{-3}$, $8\times 10^{-3}$, $2\times10^{-2}$, $6\times 10^{-2}$, and $2\times10^{-1}\textrm{M}_{\odot}$ where only the first four are nonzero in the right-hand plot.
}
\label{fig:params_m1m2}
\end{figure*}

Figure \ref{fig:params_m1m2} shows the distribution of ejecta masses using the SLy EOS, illustrating how the ejecta mass tends to scale with the component mass distribution. Among the EOS tested, SLy is nearer the lower side of ejecta distributions in both the estimated median and maximum ejecta. The fits themselves imply an ejecta distribution strongly dependent on the mass of the primary ($m_1$) and the difference between the primary and secondary masses. However, applying the fit uncertainty smears the ejecta distribution over the difference of the component masses. This effect is most evident in the marginal distributions plotted as histograms on the sides of the Figure \ref{fig:params_m1m2} panels. Since the high spin distribution has more posterior samples away from equal mass systems, as well as larger primary masses overall, more samples give rise to larger ejecta masses. While this only affects the high spin case, those EOS which allow for larger maximum masses also allow for a larger maximum ejecta values, typically $\mej > 10^{-1}$ M${}_{\odot}$ (above the maximum ejecta mass of $6.5\times10^{-2} \text{M}_{\odot}$ in the simulations to which the \mej\ fit has been calibrated). This is a natural consequence of larger maximum masses corresponding to larger differences between $m_1$ and $m_2$, as illustrated in Figure~\ref{fig:params_m1m2}.

\section{Kilonova Light Curve Models}
\label{sec:modelling}
Current kilonova emission models~\citep{LiPa1998,Me2017,BaKa2016,Tanaka2017} produce spectral energy distributions between the ultraviolet (UV) and the near-infrared (NIR). Generally, there are two different physical processes that require modeling. First, the conversion of dynamical and wind ejecta material into r-process elements (i.e., the nucleosynthesis)~\citep{KaBa2013,KaFe2015,BaKa2016,RoFe2017,Me2017}, and second, the production of an associated electromagnetic transient~\citep{MeMa2010,KaBa2013,BaKa2016,RoFe2017}. Beyond these considerations, there are still several important nuclear physics ingredients that are unknown, such as opacity and heating rate,
and can lead to large uncertainties in light curve prediction (see, e.g., \citealt{RoFe2017}). We do not attempt to model these uncertainties.

We briefly describe here three parameterized models used to generate light curves in this work. \cite{WoKo2017} use radiative transfer simulations and provide analytic fits for the peak time, bolometric luminosity, and color corrections as a function of ejecta parameters. The \cite{WoKo2017} lightcurves are scaled as a function of ejecta mass and velocity, which changes both the time of peak luminosity as well as peak magnitude. We obtain the velocity from additional fits in DU17, and assume an opacity of 10 cm${}^2$/g thus modeling the presences of lanthanides. Conversely, \cite{Me2017} provides a toy model for blue kilonova with opacity $0.1$ cm${}^2$/g for lanthanide-free matter.
DU17 use the radiative Monte-Carlo (MC) simulations of \cite{TaHo2013} and derive an analytical model for kilonova emission driven by dynamical ejecta from a BNS merger.
No wind contribution is included in DU17 although winds can potentially dominate~\citep{KiSe2015,CiKa2017,SiMe2017}.
The dynamical ejecta models tend to predict redder and more slowly rising NIR than wind-driven models.

Light curves from dynamical ejecta models depend significantly on the thermalization efficiency, the radiation transport simulations used, and other assumptions \citep{MeFe2014,CoDi2017,RoFe2017}. In our analysis we do not consider observational error from extinction in the light curve prediction, as it is likely smaller than the systematic error of the models \citep{KaKy2016}.

\section{Predicted Kilonova Light Curves}
\label{sec:lightcurves}
\begin{figure}[ht]\centering
\includegraphics[width=3.5in,height=5in]{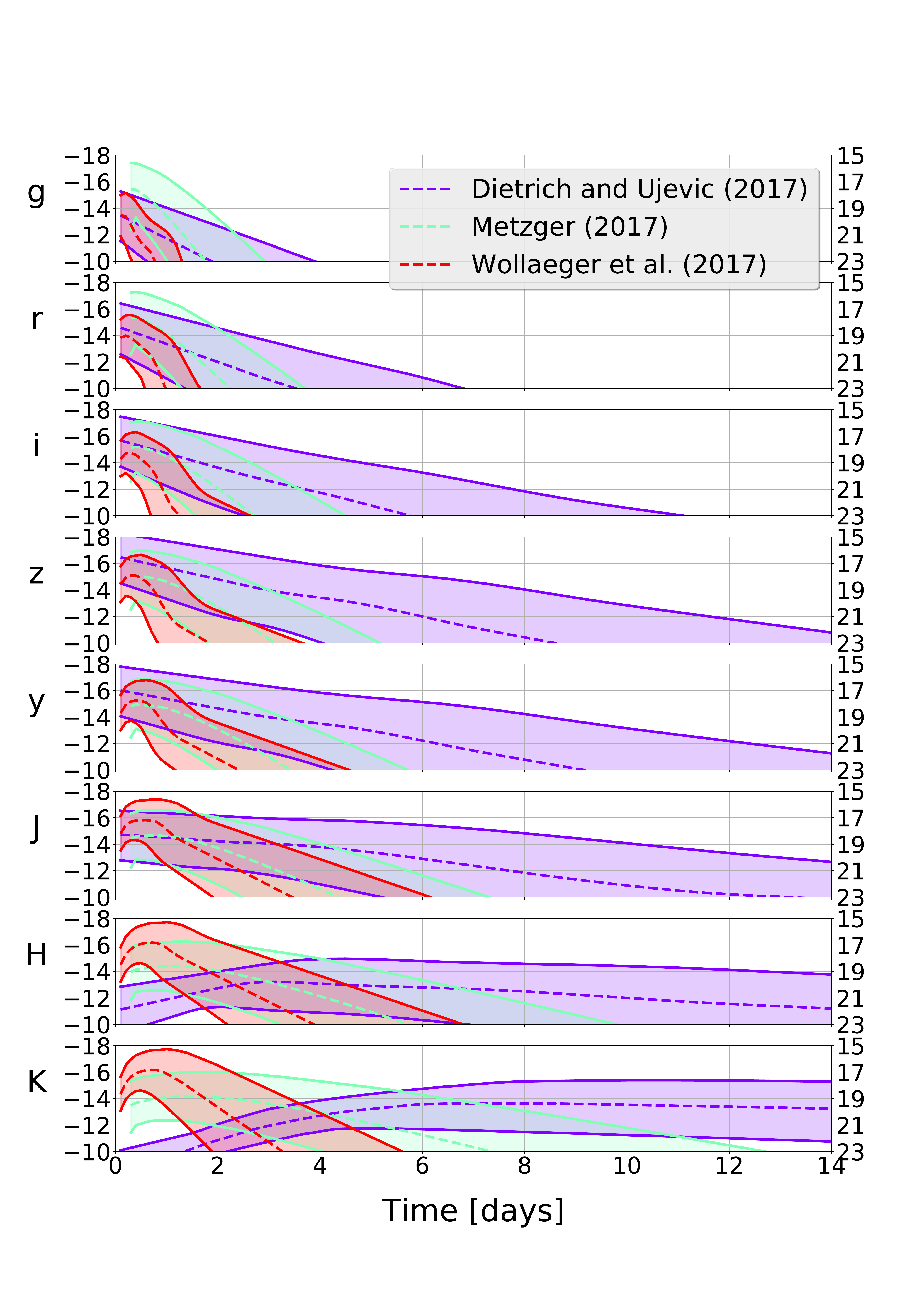}
\caption{Absolute (left vertical axis) and apparent (right vertical axis) magnitudes of light curves consistent with parameter estimation for astrophysical spins for the kilonova models of DU17; \cite{Me2017,WoKo2017} in $grizyJHK$ filters. In particular, the {\tt DZ2} model is employed from \cite{WoKo2017}. The dashed lines show the median light curve, while the shaded intervals show the 90\% intervals. In addition to including the average relative error (72\%) of the ejecta mass fitting formula, we include 1\,mag errors on the intervals to account for errors in the models themselves \citep{CoDi2017}. The lower percentiles are not conservative as we cannot definitively exclude zero ejecta mass due to unmodeled systematics. The fiducial distance to the event is $40\,{\rm Mpc}$.
}
\label{fig:mag_panels}
\end{figure}

In conjunction with the mass and tidal estimates for the low spin case,
 we calculate the mass and velocity of dynamical ejecta as described in Section~\ref{sec:kN_ejecta}. 
 Using the light curve models of DU17; \cite{Me2017,WoKo2017}, we show the absolute and apparent magnitudes consistent with these estimates of dynamical ejecta in Figure~\ref{fig:mag_panels}. 
Here, we employ the {\tt DZ2} model from \cite{WoKo2017}, and set $40\,{\rm Mpc}$ (near the median of the GW distance posterior~\citep{LVC_only_paper,H0paper}) as the fiducial distance to the event for calculating the apparent magnitudes.
DU17 exhibits the features of most lanthanide-rich dynamical ejecta models, with a rapid fade in the blue and a late rise in the NIR.
\cite{WoKo2017}, which also considers the contribution from the wind ejecta of $0.005\,M_\odot$,
 is both brighter, has a slower fade in the blue, and a faster fade in the NIR. The model in 
\cite{Me2017} --- adopted here only considering dynamical ejecta --- is between these two models, originally brighter in the blue and NIR bands (g,r,i,z) than either of these models, but fades more quickly than \cite{WoKo2017}.

\begin{figure}[ht]\centering
\includegraphics[width=\linewidth]{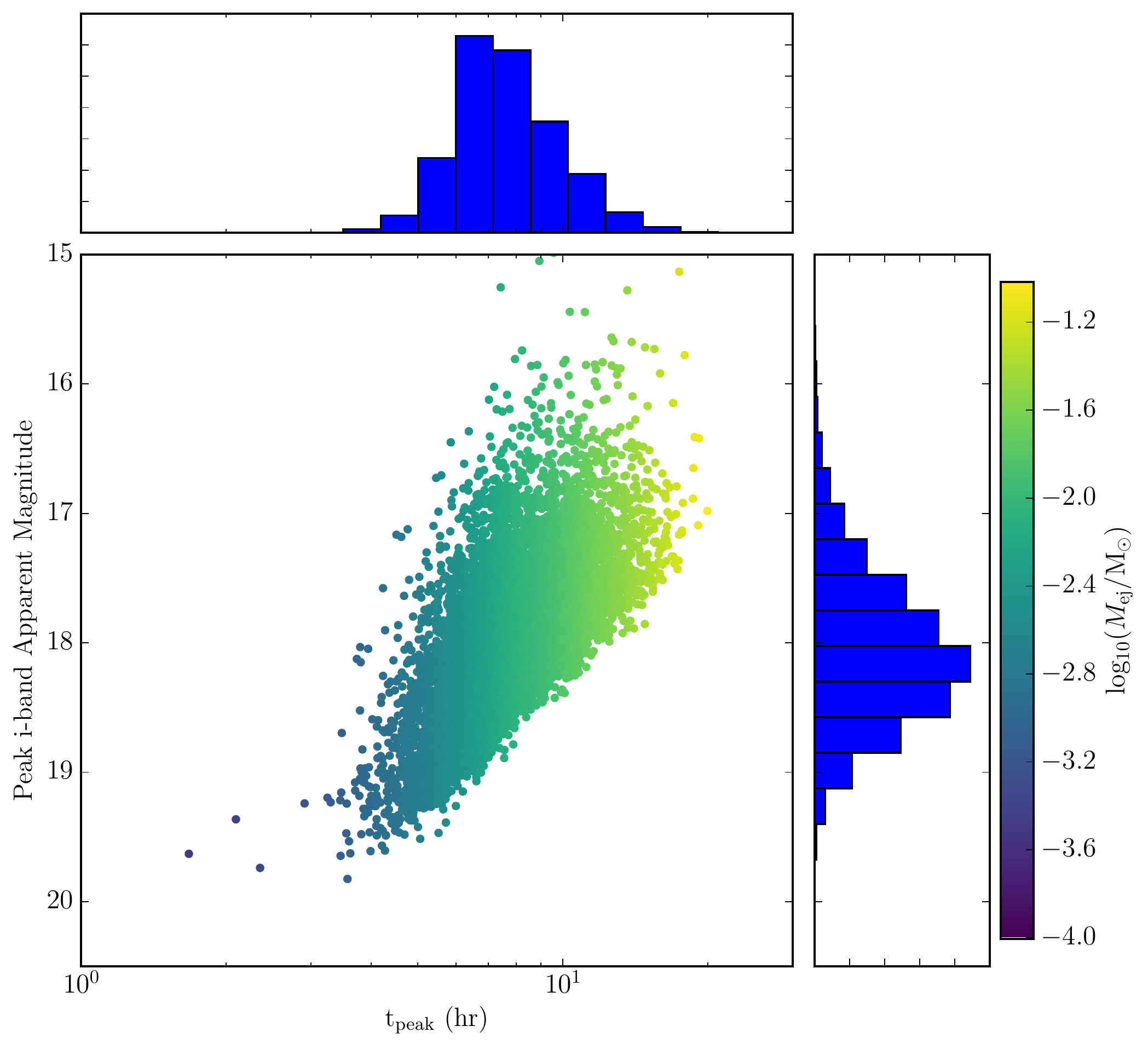}
\caption{Inferred peak i-band apparent magnitude vs. time of peak i-band magnitude with the blue model in~\cite{Me2017} and low-spin sample distribution (marginal distributions on \mej and time of peak shown on top and right). Apparent magnitudes are calculated from the dynamical ejecta only, using the GW inferred distance.}
\label{fig:peak_mag}
\end{figure}

Employing the lower opacity blue-peaked model in \cite{Me2017} and GW inferred distance, we can calculate the distribution of peak times and observed peak magnitudes in a given photometric band. Since the source resides at a low redshift, we neglect the cosmological redshift of the source. Figure \ref{fig:peak_mag} shows the peak i-band magnitudes from those light curves versus the time of peak i-band magnitude when considering the low spin distribution. The samples from the high spin distribution produce the peaks which are brighter by one magnitude on average. This is understood from the ejecta distributions in Figure \ref{fig:params_m1m2} --- the low spin distribution tends to produce less ejecta and hence is less luminous. We note again that the light curves in Figure \ref{fig:mag_panels} are calculated with a distance fixed to the source, while the magnitudes in Figure \ref{fig:peak_mag} fold in the distance inferred from the GW data. 
Thus, a wider spread arises from the variance in the GW-only distance posterior distribution. Including the distance values from the GW posteriors better estimates the variation that would arise in a prediction from only GW information as opposed to having constraints from electromagnetic measurements.

The estimates presented here are a proof-of-principle study to illustrate what is possible at present with forward modeling from GW observations.  Particularly if available before EM observations begin, or in a situation before a confident counterpart has been identified (e.g., due to poor sky localization), analysis driven by the GW data can inform EM followup observations and interpretation, particularly in cases where (due to geometric effects and observational delays) the dynamical ejecta’s effect on the light curve is enhanced. Predictions of peak times in the the emission and the color evolution are useful for comparison with early observations, and provide falsifiable predictions to evaluate models of the source.

\section{Abundance of r-Process Material}
\label{sec:rpoc_abun}
The r-process and s-process are the two known mechanisms by which heavy elements can be synthesized \citep{BuBu1957}. To assess the contribution of the r-process to the observed abundances of heavy elements \citep{2007PhR...450...97A,2008ARA&A..46..241S}, one can identify the abundances expected from the s-process alone, and hence the r-process residual. Type II SNe can produce r-process elements, but they may not produce the observed abundance patterns ~\citep[e.g.,][]{FreiburghausRembges99}.  
BNS mergers could also account for these elements.   However, quantifying the contribution of those mergers has remained elusive due to poor constraints on both the rate of mergers as well as the amount of matter ejected in each merger. With GW170817, we are able to constrain both of these quantities significantly from data. 

\begin{figure*}
\includegraphics[width=\columnwidth]{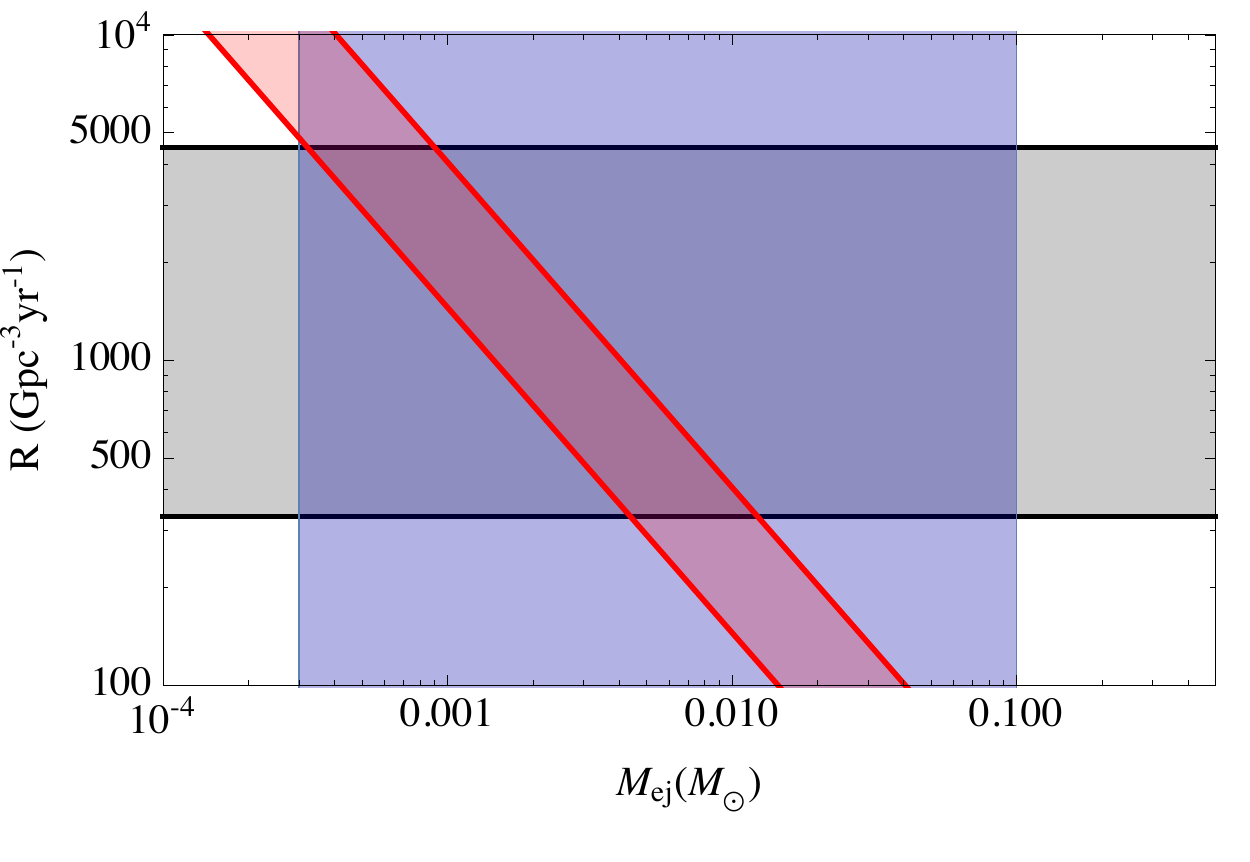}
\includegraphics[width=\columnwidth]{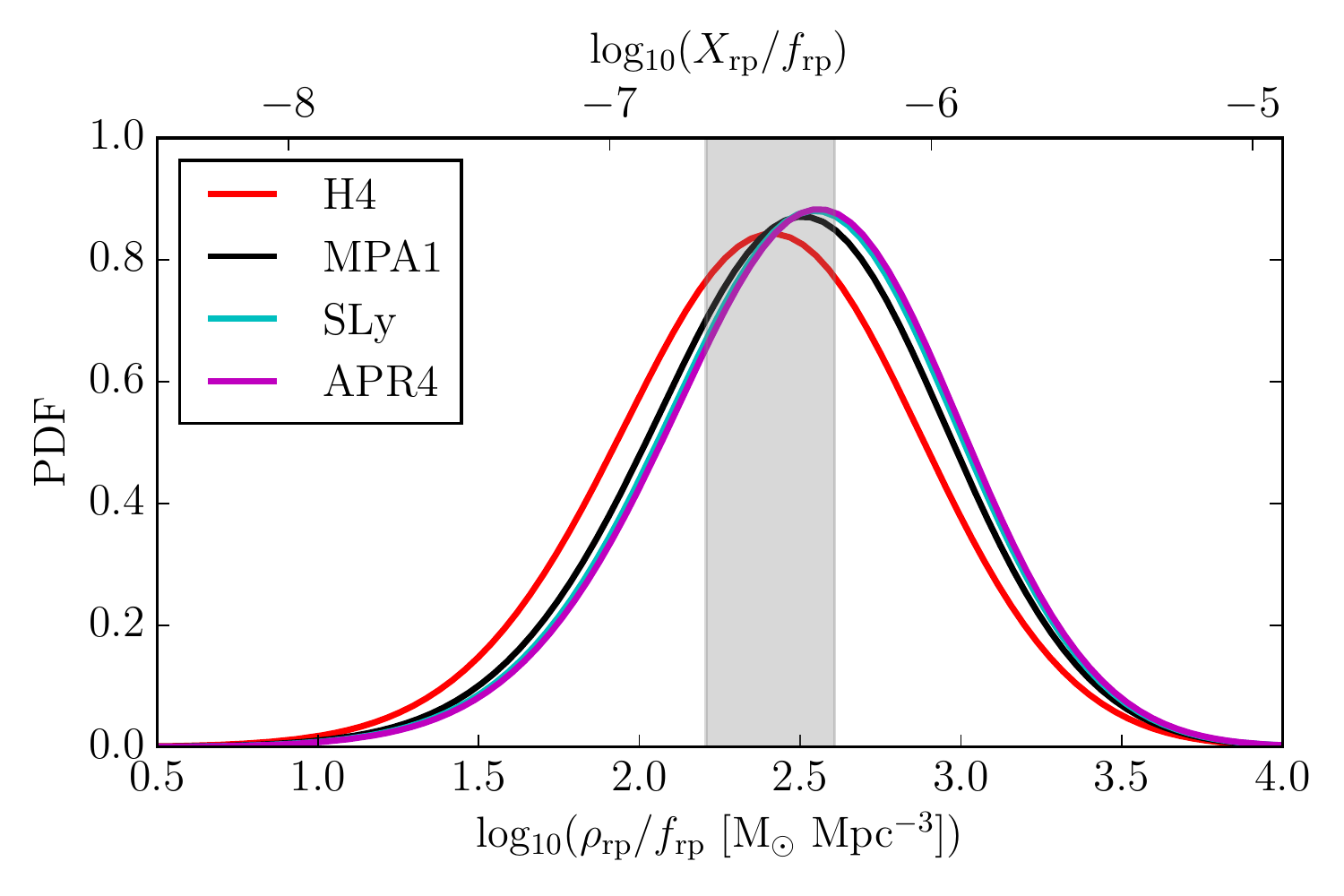}
\caption{ \label{fig:r_process_constraints} \emph{Left panel}: Plot of  the present-day BNS merger rate density $\mathcal{R}$ versus dynamical ejecta masses $M_{\rm ej}$.
The solid gray band  corresponds to the event rate range deduced from GW170817.
The solid blue band shows the approximate range of conceivable dynamical ejecta masses, based on the ejecta models used in this
work.
The red band shows the approximate range of r-process elements per unit volume, based on Galactic observations, an
approximate density of MW-like galaxies $(0.01~{\rm Mpc}^{-3}$), a range of Galactic masses, and r-process formation efficiencies $f_{\rm rp}$ between
0.5 and 1.  Configurations in the intersection of all three bands correspond to cases where dynamical ejecta from BNS mergers are solely responsible for r-process element formation.
\emph{Right panel}: Probability distributions of r-process material density and abundance (normalized by $f_{\rm rp}$) from dynamical ejecta for different EOS at $z=0$. The lower (upper) bound on the 90$\%$ credible interval for $\rho_{\rm rp}/f_{\rm rp}$ over all EOS is \MinLLrho{} ${\rm M}_\odot {\rm Mpc}^{-3}$ (\MaxULrho ${\rm M}_\odot {\rm Mpc}^{-3}$).  The
vertical gray band shows the Solar r-process abundance \citep{2007PhR...450...97A}.
}
\end{figure*}

If BNS mergers are to produce most of the observed r-process elements in the Milky Way (MW), the mergers must occur with a sufficiently high rate and eject significant amounts of r-process material.   Assuming dynamical ejecta dominate over winds, the mass fraction $X_{\rm rp}$ of r-process nuclei in the MW should be proportional  to the merger rate density $\mathcal{R}$ and dynamical ejecta mass \mej, with a proportionality constant set by the local galaxy density and the MW age and mass.
Following \cite{Qian200}, we estimate that the merger rate and ejecta per event are approximately related by $\mathcal{R}\simeq 600 (f_{\rm rp}\mej/10^{-2}\mathrm{M}_\odot)^{-1} \unit{Gpc}^{-3}\unit{yr}^{-1}$.  In this relationship, $f_{\rm rp}\equiv M_{\rm rp}/M_{\rm ej}$ is the fraction of matter dynamically ejected in NS mergers that is converted to heavy r-process elements rather than lighter products, e.g., $\alpha$ particles. The value of $f_{\rm rp}$ depends on details of the dynamics, geometry, and neutrino illumination of the ejected matter, all of which change the electron fraction ($Y_e$) distribution of ejected matter \citep[see, e.g.,][]{KaFe2015,2015MNRAS.452.3894G}. However, various studies have suggested significant r-processing of ejecta material \cite[e.g.,][]{GorielyBauswein11,2015MNRAS.452.3894G,WanajoSekiguchi14,JustBauswein15,Radice:2016dwd}.  The red band in the left panel of Figure \ref{fig:r_process_constraints} shows this relationship between  $\mathcal{R}$ and $\mej$ for $f_{\rm rp} \in [0.5,1]$ \citep[e.g.,][]{2015MNRAS.452.3894G}. Also shown in the left panel are the constraints on the local rate density of BNS mergers from GW170817 (gray) and the range of ejecta masses typically considered in the literature (blue). The overlap of these constraints suggests that BNS mergers could account for all of the observed r-process abundance.  

A more detailed calculation of r-process enrichment from the dynamical ejecta of BNS mergers can be done using the specific distributions of $\mej$ and $\mathcal{R}$ inferred from GW170817. Under the assumption that all binary mergers have the same ejecta mass as that inferred from GW170817, we calculate the average dynamically ejected local r-process material density according to 
\begin{equation}
\rho_\mathrm{rp} = f_{\rm rp}M_\mathrm{ej}\mathcal{R}
\frac{\int_0^{t_h}\int_0^t \dot{\rho}_*(\tau)p_\mathrm{delay}(t-\tau) d\tau dt}{\int_0^{t_h} \dot{\rho}_*(\tau) p_\mathrm{delay}(t_h-\tau) d\tau},
\end{equation}
where $t_h$ is the Hubble time.\footnote{We assume $\Lambda$CDM cosmology with TT+lowP+lensing+ext parameters from
  \cite{2016A&A...594A..13P}.
}   

In this expression, $\dot{\rho}_*$ is the cosmological star formation rate, assumed to follow \cite{MadauDickinson14}; 
$p_{\rm delay}$ is the delay time distribution of NS mergers, $p_\mathrm{delay}(t) \propto t^{-1}$ \citep[see,
  e.g.,][]{OShaughnBelczynski08,DominikBelczynskiI}, with minimum delay time of 10 Myr; and ${\cal R}$ is the
present-day merger
rate density for NS mergers. The denominator is a normalization factor which scales the present-day merger rate density to $\mathcal{R}$.

In the right panel of Figure \ref{fig:r_process_constraints}, we plot the distribution of $\rho_\mathrm{rp}/f_{\rm rp}$ for a few representative EOS using our $\mej$ distributions and the rates inferred from GW170817.  On the top axis, we also show $X_\mathrm{rp}/f_{\rm rp} = (\rho_\mathrm{rp} / f_{\rm rp}) / \rho_{*}$, where $\rho_{*} = \int_0^{t_h} \dot{\rho}_*(t) dt$.  If $f_{\rm rp} = 1$, the range $\MinLLrho{}  {\rm M}_\odot {\rm Mpc}^{-3} - \MaxULrho{} {\rm M}_\odot {\rm Mpc}^{-3}$ brackets our 90\% credible intervals on $\rho_{\rm rp}$ for all EoS. Both $\rho_{\rm rp}$ and $X_{\rm rp}$ are shown normalized to $f_{\rm rp}$, since $f_{\rm rp}$ depends on unknown details of the merger. The gray band in the right panel of Figure \ref{fig:r_process_constraints} shows the MW mass abundance of r-process elements, derived from \cite{2007PhR...450...97A}. As long as $f_{\rm rp} \gtrsim \frpmin$ of the dynamically ejected mass is converted to heavy r-process elements, dynamical ejecta could account for all of the MW r-process abundance.   
We have not factored in many modeling details such as the relative abundance pattern of r-process elements, the value of $f_{\rm rp}$, the relative contribution of dynamical versus wind ejecta, and uncertainties in star formation history of the Universe.  

\section{Conclusions}
\label{sec:concl}
In this Letter, we derived estimates for the dynamical ejecta mass produced by the BNS merger GW170817, as well as the corresponding kilonovae light curves and r-process nucleosynthesis yields, without additional photometric or EM spectral data. These estimates have the GW data as their foundation and use a fit to a wide variety of simulations to obtain dynamical ejecta masses from these data. Our predictions for light curves include a range of possible magnitudes and time scales of emission. In general, for the blue model in \cite{Me2017} in the i-band, 
we predict peak magnitudes concentrated between $\sim19$ and $\sim17$ for a merger consistent with our low spin results, and peak magnitude between $\sim19$ and $\sim16$ --- typically lasting twice as long --- for mergers consistent with high spin results. Such predictions can guide expectation as to whether or not future, perhaps more distant, counterparts would be observable with a given facility.
The predictions from the GW inference for the dynamically unbound matter depend strongly on the allowed spin configurations in the GW model, which in turn influence the predicted light curves. The low spin results predict smaller ejecta masses on the whole, and as such, a bright kilonova event (e.g., $> 16$ magnitude) may indicate a faster spinning NS component. We stress that the phenomenological fits used to predict \mej\ themselves are not corrected for spin effects, so  this increased brightness occurs because of degeneracies in the GW parameter estimates between spin and mass ratio.

We have also presented predicted light curves derived from other models in the literature. Our results show that when large amounts of ejecta mass are allowed, the light curves have brighter peaks and are longer-lived. They differ in color evolution, however (compare DU17 and \ \cite{WoKo2017} for example) and electromagnetic observations combined with these curves could hint towards mixtures of different ejecta material compositions~\citep{Me2017}. For example, strong emission observed in both blue and red bands could imply sectors of material containing both high and low electron fractions. However, the Metzger model, as implemented here, neglects post-merger wind effects, and in general, these conclusions only hold under the assumption that dynamical ejecta dominate the mass ejection.

Our results suggest that dynamical ejecta from rare NS mergers could be an important and inhomogeneous source of r-process elements in
the galaxy \citep{JiFrebel16,2016ApJ...832..149B}.
If more than $f_{\rm rp}\gtrsim \frpmin$ of the mass ejected from mergers is converted to r-process elements, our prediction for average r-process density in the local universe is consistent with the Galactic abundance.
Our approach does not address the contribution from  winds,  which could eject a substantial overall mass but may \citep{SiMe2017} or may not \citep{RoFe2017} have the wide range of $Y_e$ needed to produce all r-process abundances (i.e., the second and third r-process peak).
Our approach is also not as detailed as full multi-species chemical enrichment calculations used to interpret observations of individual elements in targeted populations \citep[see, e.g.,][]{2017ApJ...836..230C}. As Advanced LIGO and Virgo approach design sensitivity, these observational constraints should rapidly shrink, enabling more precise tests of the BNS r-process nucleosynthesis paradigm. Additionally, present and future electromagnetic observations should provide complementary information to directly constrain those parameters that our analysis cannot.

Finally, if electromagnetic measurements are consistent with a total ejecta mass (dynamical and wind) of $\gtrsim 0.01 M_\odot$, and if we require consistency with low neutron star spins, then one possible conclusion is that winds contribute significantly to the total ejected mass. However, if winds dominate, then the dynamical ejecta mass will be an important but potentially difficult to measure component in the light curve, which our calculations can supply. Additionally, with so much material ejected per event, to be consistent with our inferred detection rate, we would predict that only a fraction of the ejecta can form r-process elements.

The coincidence of GW170817 and GRB170817A was an exceptionally rare event, allowing for a unique set of measurements to be made about the processes driven by the BNS merger. Future observations should facilitate the refinement of these measurements. The observation of GW170817 suggests that in the upcoming year-long third observing run~\citep{lrr-2016-1} with a three instrument GW network, there will likely be more GW observations of BNS. In the coming years, GW measurements will allow for better understanding of populations of kilonova events.

\acknowledgements
The authors gratefully acknowledge the support of the United States
National Science Foundation (NSF) for the construction and operation of the
LIGO Laboratory and Advanced LIGO as well as the Science and Technology Facilities Council (STFC) of the
United Kingdom, the Max-Planck-Society (MPS), and the State of
Niedersachsen/Germany for support of the construction of Advanced LIGO 
and construction and operation of the GEO600 detector. 
Additional support for Advanced LIGO was provided by the Australian Research Council.
The authors gratefully acknowledge the Italian Istituto Nazionale di Fisica Nucleare (INFN),  
the French Centre National de la Recherche Scientifique (CNRS) and
the Foundation for Fundamental Research on Matter supported by the Netherlands Organisation for Scientific Research, 
for the construction and operation of the Virgo detector
and the creation and support  of the EGO consortium. 
The authors also gratefully acknowledge research support from these agencies as well as by 
the Council of Scientific and Industrial Research of India, 
the Department of Science and Technology, India,
the Science \& Engineering Research Board (SERB), India,
the Ministry of Human Resource Development, India,
the Spanish  Agencia Estatal de Investigaci\'on,
the  Vicepresid\`encia i Conselleria d'Innovaci\'o, Recerca i Turisme and the Conselleria d'Educaci\'o i Universitat del Govern de les Illes Balears,
the Conselleria d'Educaci\'o, Investigaci\'o, Cultura i Esport de la Generalitat Valenciana,
the National Science Centre of Poland,
the Swiss National Science Foundation (SNSF),
the Russian Foundation for Basic Research, 
the Russian Science Foundation,
the European Commission,
the European Regional Development Funds (ERDF),
the Royal Society, 
the Scottish Funding Council, 
the Scottish Universities Physics Alliance, 
the Hungarian Scientific Research Fund (OTKA),
the Lyon Institute of Origins (LIO),
the National Research, Development and Innovation Office Hungary (NKFI), 
the National Research Foundation of Korea,
Industry Canada and the Province of Ontario through the Ministry of Economic Development and Innovation, 
the Natural Science and Engineering Research Council Canada,
the Canadian Institute for Advanced Research,
the Brazilian Ministry of Science, Technology, Innovations, and Communications,
the International Center for Theoretical Physics South American Institute for Fundamental Research (ICTP-SAIFR), 
the Research Grants Council of Hong Kong,
the National Natural Science Foundation of China (NSFC),
the Leverhulme Trust, 
the Research Corporation, 
the Ministry of Science and Technology (MOST), Taiwan
and
the Kavli Foundation.
The authors gratefully acknowledge the support of the NSF, STFC, MPS, INFN, CNRS and the
State of Niedersachsen/Germany for provision of computational resources.

\bibliographystyle{yahapj}
\bibliography{references}

\begin{thebibliography}{}
\providecommand\natexlab[1]{#1}
\providecommand\JournalTitle[1]{#1}

\bibitem[{Abbott {et~al.}(2016{\natexlab{a}})Abbott, Abbott, Abbott,
  {et~al.}}]{GW150914PE}
Abbott, B.~P., Abbott, R., Abbott, T.~D., {et~al.} 2016{\natexlab{a}},
  \href{http://dx.doi.org/10.1103/PhysRevLett.116.241102}{\JournalTitle{Phys.
  Rev. Lett.}, 116, 241102}

\bibitem[{Abbott {et~al.}(2016{\natexlab{b}})Abbott, Abbott, Abbott,
  {et~al.}}]{lrr-2016-1}
---. 2016{\natexlab{b}},
  \href{http://dx.doi.org/10.1007/lrr-2016-1}{\JournalTitle{LRR}, 19, 1}

\bibitem[{Abbott {et~al.}(2017{\natexlab{a}})Abbott, Abbott, Abbott,
  {et~al.}}]{H0paper}
---. 2017{\natexlab{a}},
  \href{http://dx.doi.org/http://doi.org/10.1038/nature24471}{\JournalTitle{Nature},
  in press}

\bibitem[{Abbott {et~al.}(2017{\natexlab{b}})Abbott, Abbott, Abbott,
  {et~al.}}]{GRBpaper}
---. 2017{\natexlab{b}},
  \href{http://dx.doi.org/10.3847/2041-8213/aa920c}{\JournalTitle{ApJL}, in
  press}

\bibitem[{Abbott {et~al.}(2017{\natexlab{c}})Abbott, Abbott, Abbott,
  {et~al.}}]{LVC_only_paper}
---. 2017{\natexlab{c}},
  \href{http://dx.doi.org/10.1103/PhysRevLett.119.161101}{\JournalTitle{\prl},
  119, 61101}

\bibitem[{Abbott {et~al.}(2017{\natexlab{d}})Abbott, Abbott, Abbott,
  {et~al.}}]{MMAPaper}
---. 2017{\natexlab{d}},
  \href{http://dx.doi.org/10.3847/2041-8213/aa91c9}{\JournalTitle{ApJL}, in
  press}

\bibitem[{{Ade} {et~al.}(2016){Ade}, {Aghanim}, {Arnaud}, {Ashdown}, {Aumont},
  {Baccigalupi}, {Banday}, {Barreiro}, {Bartlett}, \&
  et~al.}]{2016A&A...594A..13P}
{Ade}, P.~A.~R., {Aghanim}, N., {Arnaud}, M., {et~al.} 2016,
  \href{http://dx.doi.org/10.1051/0004-6361/201525830}{\JournalTitle{\aap},
  594, A13}

\bibitem[{Akmal {et~al.}(1998)Akmal, Pandharipande, \& Ravenhall}]{AkPa1998}
Akmal, A., Pandharipande, V.~R., \& Ravenhall, D.~G. 1998,
  \href{http://dx.doi.org/10.1103/PhysRevC.58.1804}{\JournalTitle{Phys. Rev.
  C}, 58, 1804}

\bibitem[{Alexander {et~al.}(2017)}]{Alexander17}
Alexander, K.~D., {et~al.} 2017,
  \href{http://dx.doi.org/10.3847/2041-8213/aa905d}{\JournalTitle{ApJL}, in
  press}

\bibitem[{Antoniadis {et~al.}(2013)Antoniadis, Freire, Wex,
  {et~al.}}]{AnFr2013}
Antoniadis, J., Freire, P. C.~C., Wex, N., {et~al.} 2013,
  \href{http://dx.doi.org/10.1126/science.1233232}{\JournalTitle{Science}, 340,
  6131}

\bibitem[{{Arcavi} {et~al.}(2017){Arcavi}, Hosseinzadeh, Howell,
  {et~al.}}]{LASCUMBRES}
{Arcavi}, I., Hosseinzadeh, G., Howell, D.~A., {et~al.} 2017,
  \href{http://dx.doi.org/10.1038/nature24291}{\JournalTitle{Nature}, in press}

\bibitem[{{Arnould} {et~al.}(2007){Arnould}, {Goriely}, \&
  {Takahashi}}]{2007PhR...450...97A}
{Arnould}, M., {Goriely}, S., \& {Takahashi}, K. 2007,
  \href{http://dx.doi.org/10.1016/j.physrep.2007.06.002}{\JournalTitle{\physrep},
  450, 97}

\bibitem[{Barnes \& Kasen(2013)}]{Baka2013}
Barnes, J., \& Kasen, D. 2013,
  \href{http://dx.doi.org/10.1088/0004-637X/775/1/18}{\JournalTitle{Astrophys.
  J.}, 775, 18}

\bibitem[{Barnes {et~al.}(2016)Barnes, Kasen, Wu, \&
  Mart\'inez-Pinedo}]{BaKa2016}
Barnes, J., Kasen, D., Wu, M.-R., \& Mart\'inez-Pinedo, G. 2016,
  \href{http://dx.doi.org/10.3847/0004-637X/829/2/110}{\JournalTitle{\apj},
  829, 110}

\bibitem[{{Bauswein} {et~al.}(2013){Bauswein}, {Goriely}, \&
  {Janka}}]{BaGo2013}
{Bauswein}, A., {Goriely}, S., \& {Janka}, H.-T. 2013,
  \href{http://dx.doi.org/10.1088/0004-637X/773/1/78}{\JournalTitle{\apj}, 773,
  78}

\bibitem[{{Beniamini} {et~al.}(2016){Beniamini}, {Hotokezaka}, \&
  {Piran}}]{2016ApJ...832..149B}
{Beniamini}, P., {Hotokezaka}, K., \& {Piran}, T. 2016,
  \href{http://dx.doi.org/10.3847/0004-637X/832/2/149}{\JournalTitle{\apj},
  832, 149}

\bibitem[{{Bovard} {et~al.}(2017){Bovard}, {Martin}, {Guercilena}, {Arcones},
  {Rezzolla}, \& {Korobkin}}]{BoMa2017}
{Bovard}, L., {Martin}, D., {Guercilena}, F., {et~al.} 2017,
  \JournalTitle{ArXiv e-prints},
  \href{http://arxiv.org/abs/1709.09630}{{\sffamily arXiv:1709.09630 [gr-qc]}}

\bibitem[{Buchdahl(1959)}]{Buc1959}
Buchdahl, H.~A. 1959,
  \href{http://dx.doi.org/10.1103/PhysRev.116.1027}{\JournalTitle{Phys. Rev.},
  116, 1027}

\bibitem[{Burbidge {et~al.}(1957)Burbidge, Burbidge, Fowler, \&
  Hoyle}]{BuBu1957}
Burbidge, E.~M., Burbidge, G.~R., Fowler, W.~A., \& Hoyle, F. 1957,
  \href{http://dx.doi.org/10.1103/RevModPhys.29.547}{\JournalTitle{Rev. Mod.
  Phys.}, 29, 547}

\bibitem[{Burgay {et~al.}(2003)Burgay, D'Amico, Possenti, Manchester, Lyne,
  Joshi, McLaughlin, Kramer, Sarkissian, Camilo, Kalogera, Kim, \&
  Lorimer}]{Bu2003}
Burgay, M., D'Amico, N., Possenti, A., {et~al.} 2003,
  \href{http://dx.doi.org/10.1038/nature02124}{\JournalTitle{Nature}, 426, 531}

\bibitem[{{Chornock} {et~al.}(2017)}]{Chornock17}
{Chornock}, R., {et~al.} 2017,
  \href{http://dx.doi.org/10.3847/2041-8213/aa905c}{\JournalTitle{ApJL}, in
  press}

\bibitem[{{Ciolfi} {et~al.}(2017){Ciolfi}, {Kastaun}, {Giacomazzo}, {Endrizzi},
  {Siegel}, \& {Perna}}]{CiKa2017}
{Ciolfi}, R., {Kastaun}, W., {Giacomazzo}, B., {et~al.} 2017,
  \href{http://dx.doi.org/10.1103/PhysRevD.95.063016}{\JournalTitle{\prd}, 95,
  063016}

\bibitem[{{C{\^o}t{\'e}} {et~al.}(2017){C{\^o}t{\'e}}, {Belczynski}, {Fryer},
  {Ritter}, {Paul}, {Wehmeyer}, \& {O'Shea}}]{2017ApJ...836..230C}
{C{\^o}t{\'e}}, B., {Belczynski}, K., {Fryer}, C.~L., {et~al.} 2017,
  \href{http://dx.doi.org/10.3847/1538-4357/aa5c8d}{\JournalTitle{\apj}, 836,
  230}

\bibitem[{{Coughlin} {et~al.}(2017){Coughlin}, {Dietrich}, {Kawaguchi},
  {Smartt}, {Stubbs}, \& {Ujevic}}]{CoDi2017}
{Coughlin}, M., {Dietrich}, T., {Kawaguchi}, K., {et~al.} 2017,
  \JournalTitle{ArXiv e-prints},
  \href{http://arxiv.org/abs/1708.07714}{{\sffamily arXiv:1708.07714
  [astro-ph.HE]}}

\bibitem[{Coulter {et~al.}(2017)}]{SWOPE}
Coulter, D.~A., {et~al.} 2017,
  \href{http://dx.doi.org/10.1126/science.aap9811}{\JournalTitle{Science}, in
  press}

\bibitem[{Covino {et~al.}(2017)Covino, Wiersema, Fan, {et~al.}}]{CoWi2017}
Covino, S., Wiersema, K., Fan, Y.~Z., {et~al.} 2017,
  \href{http://dx.doi.org/10.1038/s41550-017-0285-z}{\JournalTitle{Nature
  Astronomy}, in press}

\bibitem[{{Cowperthwaite} {et~al.}(2017)}]{Cowperthwaite17}
{Cowperthwaite}, P., {et~al.} 2017,
  \href{http://dx.doi.org/10.3847/2041-8213/aa8fc7}{\JournalTitle{ApJL}, in
  press}

\bibitem[{Damour {et~al.}(2012)Damour, Nagar, \& Villain}]{DaNa2012}
Damour, T., Nagar, A., \& Villain, L. 2012,
  \href{http://dx.doi.org/10.1103/PhysRevD.85.123007}{\JournalTitle{\prd}, 85,
  123007}

\bibitem[{Del~Pozzo {et~al.}(2013)Del~Pozzo, Li, Agathos, Van Den~Broeck, \&
  Vitale}]{DeLi2013}
Del~Pozzo, W., Li, T. G.~F., Agathos, M., Van Den~Broeck, C., \& Vitale, S.
  2013,
  \href{http://dx.doi.org/10.1103/PhysRevLett.111.071101}{\JournalTitle{Phys.
  Rev. Lett.}, 111, 071101}

\bibitem[{Demorest {et~al.}(2010)Demorest, Pennucci, Ransom, Roberts, \&
  Hessels}]{DePe2010}
Demorest, P., Pennucci, T., Ransom, S., Roberts, M., \& Hessels, J. 2010,
  \href{http://dx.doi.org/10.1038/nature09466}{\JournalTitle{Nature}, 467,
  1081}

\bibitem[{Dessart {et~al.}(2009)Dessart, Ott, Burrows, Rosswog, \&
  Livne}]{DeOt2008}
Dessart, L., Ott, C., Burrows, A., Rosswog, S., \& Livne, E. 2009,
  \href{http://dx.doi.org/10.1088/0004-637X/690/2/1681}{\JournalTitle{\apj},
  690, 1681}

\bibitem[{{Dessart} {et~al.}(2009){Dessart}, {Ott}, {Burrows}, {Rosswog}, \&
  {Livne}}]{DeOt2009}
{Dessart}, L., {Ott}, C.~D., {Burrows}, A., {Rosswog}, S., \& {Livne}, E. 2009,
  \href{http://dx.doi.org/10.1088/0004-637X/690/2/1681}{\JournalTitle{\apj},
  690, 1681}

\bibitem[{Diaz {et~al.}(2017)Diaz, Macri, {et~al.}}]{DiMa2017}
Diaz, M.~C., Macri, L.~M., {et~al.} 2017,
  \href{http://dx.doi.org/10.3847/2041-8213/aa9060}{\JournalTitle{ApJL}, in
  press}

\bibitem[{Dietrich {et~al.}(2017{\natexlab{a}})Dietrich, Bernuzzi, Ujevic, \&
  Tichy}]{DiBe2017}
Dietrich, T., Bernuzzi, S., Ujevic, M., \& Tichy, W. 2017{\natexlab{a}},
  \href{http://dx.doi.org/10.1103/PhysRevD.95.044045}{\JournalTitle{Phys. Rev.
  D}, 95, 044045}

\bibitem[{Dietrich \& Ujevic(2017)}]{DiUj2017}
Dietrich, T., \& Ujevic, M. 2017,
  \href{http://dx.doi.org/10.1088/1361-6382/aa6bb0}{\JournalTitle{Class. Quant.
  Grav.}, 34, 105014}

\bibitem[{Dietrich {et~al.}(2017{\natexlab{b}})Dietrich, Ujevic, Tichy,
  Bernuzzi, \& Br{\"u}gmann}]{Dietrich:2016hky}
Dietrich, T., Ujevic, M., Tichy, W., Bernuzzi, S., \& Br{\"u}gmann, B.
  2017{\natexlab{b}},
  \href{http://dx.doi.org/10.1103/PhysRevD.95.024029}{\JournalTitle{\prd}, 95,
  024029}

\bibitem[{{Dominik} {et~al.}(2012){Dominik}, {Belczynski}, {Fryer}, {Holz},
  {Berti}, {Bulik}, {Mandel}, \& {O'Shaughnessy}}]{DominikBelczynskiI}
{Dominik}, M., {Belczynski}, K., {Fryer}, C., {et~al.} 2012,
  \href{http://dx.doi.org/10.1088/0004-637X/759/1/52}{\JournalTitle{\apj}, 759,
  52}

\bibitem[{Douchin \& Haensel(2001)}]{DoHa2001}
Douchin, F., \& Haensel, P. 2001,
  \href{http://dx.doi.org/10.1051/0004-6361:20011402}{\JournalTitle{\aap}, 380,
  151}

\bibitem[{{Drout} {et~al.}(2017)}]{MAGELLAN}
{Drout}, M.~R., {et~al.} 2017,
  \href{http://dx.doi.org/10.1126/science.aaq0049}{\JournalTitle{Science}, in
  press}

\bibitem[{{Endrizzi} {et~al.}(2016){Endrizzi}, {Ciolfi}, {Giacomazzo},
  {Kastaun}, \& {Kawamura}}]{EnCi2016}
{Endrizzi}, A., {Ciolfi}, R., {Giacomazzo}, B., {Kastaun}, W., \& {Kawamura},
  T. 2016,
  \href{http://dx.doi.org/10.1088/0264-9381/33/16/164001}{\JournalTitle{Class.
  Quant. Grav.}, 33, 164001}

\bibitem[{Engvik {et~al.}(1996)Engvik, Bao, Hjorth-Jensen, Osnes, \&
  Ostgaard}]{EnBa1996}
Engvik, L., Bao, G., Hjorth-Jensen, M., Osnes, E., \& Ostgaard, E. 1996,
  \href{http://dx.doi.org/10.1086/177827}{\JournalTitle{\apj}, 469, 794}

\bibitem[{{Evans} {et~al.}(2017)}]{SWIFT}
{Evans}, P.~A., {et~al.} 2017,
  \href{http://dx.doi.org/10.1126/science.aap9580}{\JournalTitle{Science}, in
  press}

\bibitem[{{Fern{\'a}ndez} {et~al.}(2015){Fern{\'a}ndez}, {Kasen}, {Metzger}, \&
  {Quataert}}]{FeKa2015}
{Fern{\'a}ndez}, R., {Kasen}, D., {Metzger}, B.~D., \& {Quataert}, E. 2015,
  \href{http://dx.doi.org/10.1093/mnras/stu2112}{\JournalTitle{\mnras}, 446,
  750}

\bibitem[{Flanagan \& Hinderer(2008)}]{FlHi2008}
Flanagan, {\'E}.~{\'E}., \& Hinderer, T. 2008,
  \href{http://dx.doi.org/10.1103/PhysRevD.77.021502}{\JournalTitle{Phys. Rev.
  D}, 77, 021502}

\bibitem[{{Foucart} {et~al.}(2016){Foucart}, {O'Connor}, {Roberts}, {Kidder},
  {Pfeiffer}, \& {Scheel}}]{FoOC2016}
{Foucart}, F., {O'Connor}, E., {Roberts}, L., {et~al.} 2016,
  \href{http://dx.doi.org/10.1103/PhysRevD.94.123016}{\JournalTitle{\prd}, 94,
  123016}

\bibitem[{Freiburghaus {et~al.}(1999)Freiburghaus, Rembges, Rauscher, Kolbe,
  Thielemann, Kratz, Pfeiffer, \& Cowan}]{FreiburghausRembges99}
Freiburghaus, C., Rembges, J.-F., Rauscher, T., {et~al.} 1999,
  \href{http://stacks.iop.org/0004-637X/516/i=1/a=381}{\JournalTitle{\apj},
  516, 381}

\bibitem[{{Fujibayashi} {et~al.}(2017){Fujibayashi}, {Sekiguchi}, {Kiuchi}, \&
  {Shibata}}]{FuSe2017}
{Fujibayashi}, S., {Sekiguchi}, Y., {Kiuchi}, K., \& {Shibata}, M. 2017,
  \JournalTitle{ArXiv e-prints},
  \href{http://arxiv.org/abs/1703.10191}{{\sffamily arXiv:1703.10191
  [astro-ph.HE]}}, {A}pJ, accepted

\bibitem[{Glendenning(1985)}]{Gl1985}
Glendenning, N.~K. 1985,
  \href{http://dx.doi.org/10.1086/163253}{\JournalTitle{\apj}, 293, 470}

\bibitem[{{Goriely} {et~al.}(2011){Goriely}, {Bauswein}, \&
  {Janka}}]{GorielyBauswein11}
{Goriely}, S., {Bauswein}, A., \& {Janka}, H.-T. 2011,
  \href{http://dx.doi.org/10.1088/2041-8205/738/2/L32}{\JournalTitle{ApJL},
  738, L32}

\bibitem[{{Goriely} {et~al.}(2015){Goriely}, {Bauswein}, {Just}, {Pllumbi}, \&
  {Janka}}]{2015MNRAS.452.3894G}
{Goriely}, S., {Bauswein}, A., {Just}, O., {Pllumbi}, E., \& {Janka}, H.-T.
  2015, \href{http://dx.doi.org/10.1093/mnras/stv1526}{\JournalTitle{\mnras},
  452, 3894}

\bibitem[{{Hotokezaka} {et~al.}(2013){Hotokezaka}, {Kiuchi}, {Kyutoku},
  {Okawa}, {Sekiguchi}, {Shibata}, \& {Taniguchi}}]{HoKi2013}
{Hotokezaka}, K., {Kiuchi}, K., {Kyutoku}, K., {et~al.} 2013,
  \href{http://dx.doi.org/10.1103/PhysRevD.87.024001}{\JournalTitle{\prd}, 87,
  024001}

\bibitem[{{Ji} {et~al.}(2016){Ji}, {Frebel}, {Chiti}, \& {Simon}}]{JiFrebel16}
{Ji}, A.~P., {Frebel}, A., {Chiti}, A., \& {Simon}, J.~D. 2016,
  \href{http://dx.doi.org/10.1038/nature17425}{\JournalTitle{\nat}, 531, 610}

\bibitem[{{Just} {et~al.}(2015){Just}, {Bauswein}, {Pulpillo}, {Goriely}, \&
  {Janka}}]{JustBauswein15}
{Just}, O., {Bauswein}, A., {Pulpillo}, R.~A., {Goriely}, S., \& {Janka}, H.-T.
  2015, \href{http://dx.doi.org/10.1093/mnras/stv009}{\JournalTitle{\mnras},
  448, 541}

\bibitem[{Kalogera \& Baym(1996)}]{KaBa1996}
Kalogera, V., \& Baym, G. 1996,
  \href{http://dx.doi.org/10.1086/310296}{\JournalTitle{ApJL}, 470, L61}

\bibitem[{Kasen {et~al.}(2013)Kasen, Badnell, \& Barnes}]{KaBa2013}
Kasen, D., Badnell, N.~R., \& Barnes, J. 2013,
  \href{http://dx.doi.org/10.1088/0004-637X/774/1/25}{\JournalTitle{\apj}, 774,
  25}

\bibitem[{Kasen {et~al.}(2015)Kasen, Fernandez, \& Metzger}]{KaFe2015}
Kasen, D., Fernandez, R., \& Metzger, B. 2015,
  \href{http://dx.doi.org/10.1093/mnras/stv721}{\JournalTitle{\mnras}, 450,
  1777}

\bibitem[{Kasen {et~al.}(2017)}]{KaEA2017}
Kasen, D., {et~al.} 2017,
  \href{http://dx.doi.org/10.1038/nature24453}{\JournalTitle{Nature}, in press}

\bibitem[{Kastaun {et~al.}(2017)Kastaun, Ciolfi, Endrizzi, \&
  Giacomazzo}]{Kastaun:2016elu}
Kastaun, W., Ciolfi, R., Endrizzi, A., \& Giacomazzo, B. 2017,
  \href{http://dx.doi.org/10.1103/PhysRevD.96.043019}{\JournalTitle{\prd}, 96,
  043019}

\bibitem[{Kastaun \& Galeazzi(2015)}]{Kastaun:2014fna}
Kastaun, W., \& Galeazzi, F. 2015,
  \href{http://dx.doi.org/10.1103/PhysRevD.91.064027}{\JournalTitle{\prd}, 91,
  064027}

\bibitem[{Kawaguchi {et~al.}(2016)Kawaguchi, Kyutoku, Shibata, \&
  Tanaka}]{KaKy2016}
Kawaguchi, K., Kyutoku, K., Shibata, M., \& Tanaka, M. 2016,
  \href{http://dx.doi.org/10.3847/0004-637X/825/1/52}{\JournalTitle{\apj}, 825,
  52}

\bibitem[{{Kiuchi} {et~al.}(2015){Kiuchi}, {Sekiguchi}, {Kyutoku}, {Shibata},
  {Taniguchi}, \& {Wada}}]{KiSe2015}
{Kiuchi}, K., {Sekiguchi}, Y., {Kyutoku}, K., {et~al.} 2015,
  \href{http://dx.doi.org/10.1103/PhysRevD.92.064034}{\JournalTitle{\prd}, 92,
  064034}

\bibitem[{Lackey {et~al.}(2006)Lackey, Nayyar, \& Owen}]{LaNa2006}
Lackey, B.~D., Nayyar, M., \& Owen, B.~J. 2006,
  \href{http://dx.doi.org/10.1103/PhysRevD.73.024021}{\JournalTitle{Phys. Rev.
  D}, 73, 024021}

\bibitem[{Lattimer \& Prakash(2001)}]{LaPr2001}
Lattimer, J.~M., \& Prakash, M. 2001,
  \href{http://dx.doi.org/10.1086/319702}{\JournalTitle{\apj}, 550, 426}

\bibitem[{{Lattimer} \& {Schramm}(1974)}]{LaSc1974}
{Lattimer}, J.~M., \& {Schramm}, D.~N. 1974,
  \href{http://dx.doi.org/10.1086/181612}{\JournalTitle{ApJL}, 192, L145}

\bibitem[{Lehner {et~al.}(2016)Lehner, Liebling, Palenzuela, Caballero,
  O'Connor, Anderson, \& Neilsen}]{LeLi2016}
Lehner, L., Liebling, S.~L., Palenzuela, C., {et~al.} 2016,
  \href{http://dx.doi.org/10.1088/0264-9381/33/18/184002}{\JournalTitle{Class.
  Quant. Grav.}, 33, 184002}

\bibitem[{Li \& Paczynski(1998)}]{LiPa1998}
Li, L.-X., \& Paczynski, B. 1998,
  \href{http://stacks.iop.org/1538-4357/507/i=1/a=L59}{\JournalTitle{ApJL},
  507, L59}

\bibitem[{Lipunov {et~al.}(2017)}]{MASTER}
Lipunov, V.~N., {et~al.} 2017,
  \href{http://dx.doi.org/10.3847/2041-8213/aa92c0}{\JournalTitle{ApJL}, in
  press}

\bibitem[{Lo \& Lin(2011)}]{LoLi2011}
Lo, K.-W., \& Lin, L.-M. 2011,
  \href{http://dx.doi.org/10.1088/0004-637X/728/1/12}{\JournalTitle{\apj}, 728,
  12}

\bibitem[{{Madau} \& {Dickinson}(2014)}]{MadauDickinson14}
{Madau}, P., \& {Dickinson}, M. 2014,
  \href{http://dx.doi.org/10.1146/annurev-astro-081811-125615}{\JournalTitle{\araa},
  52, 415}

\bibitem[{Martin {et~al.}(2015)Martin, Perego, Arcones, Thielemann, Korobkin,
  \& Rosswog}]{Martin:2015hxa}
Martin, D., Perego, A., Arcones, A., {et~al.} 2015,
  \href{http://dx.doi.org/10.1088/0004-637X/813/1/2}{\JournalTitle{\apj}, 813,
  2}

\bibitem[{{McCully} {et~al.}(2017){McCully}, {Hiramatsu}, {Howell},
  {Hosseinzadeh}, {Arcavi}, {Kasen}, {Barnes}, {Shara}, \&
  {Williams}}]{McCully2017}
{McCully}, C., {Hiramatsu}, D., {Howell}, D.~A., {et~al.} 2017,
  \href{http://dx.doi.org/10.3847/2041-8213/aa9111}{\JournalTitle{ApJL}, in
  press}

\bibitem[{Metzger(2017)}]{Me2017}
Metzger, B.~D. 2017,
  \href{http://dx.doi.org/10.1007/s41114-017-0006-z}{\JournalTitle{LRR}, 20, 3}

\bibitem[{Metzger \& Berger(2012)}]{MeBe2012}
Metzger, B.~D., \& Berger, E. 2012,
  \href{http://dx.doi.org/10.1088/0004-637X/746/1/48}{\JournalTitle{\apj}, 746,
  48}

\bibitem[{Metzger \& Fernandez(2014)}]{MeFe2014}
Metzger, B.~D., \& Fernandez, R. 2014,
  \href{http://dx.doi.org/10.1093/mnras/stu802}{\JournalTitle{\mnras}, 441,
  3444}

\bibitem[{{Metzger} {et~al.}(2010){Metzger}, {Mart{\'{\i}}nez-Pinedo},
  {Darbha}, {Quataert}, {Arcones}, {Kasen}, {Thomas}, {Nugent}, {Panov}, \&
  {Zinner}}]{MeMa2010}
{Metzger}, B.~D., {Mart{\'{\i}}nez-Pinedo}, G., {Darbha}, S., {et~al.} 2010,
  \href{http://dx.doi.org/10.1111/j.1365-2966.2010.16864.x}{\JournalTitle{\mnras},
  406, 2650}

\bibitem[{M{\"u}ller \& Serot(1996)}]{MuSe1996}
M{\"u}ller, H., \& Serot, B.~D. 1996,
  \href{http://dx.doi.org/10.1016/0375-9474(96)00187-X}{\JournalTitle{Nucl.
  Phys.}, A606, 508}

\bibitem[{M{\"u}ther {et~al.}(1987)M{\"u}ther, Prakash, \&
  Ainsworth}]{MuPr1987}
M{\"u}ther, H., Prakash, M., \& Ainsworth, T.~L. 1987,
  \href{http://dx.doi.org/10.1016/0370-2693(87)91611-X}{\JournalTitle{Phys.
  Lett.}, B199, 469}

\bibitem[{Nakar(2007)}]{Nakar2007}
Nakar, E. 2007,
  \href{http://dx.doi.org/10.1016/j.physrep.2007.02.005}{\JournalTitle{\physrep},
  442, 166}

\bibitem[{Nicholl {et~al.}(2017)}]{Nicholl17}
Nicholl, M., {et~al.} 2017,
  \href{http://dx.doi.org/10.3847/2041-8213/aa9029}{\JournalTitle{ApJL}, in
  press}

\bibitem[{Oertel {et~al.}(2017)Oertel, Hempel, Kl{\"a}hn, \& Typel}]{OeHe2017}
Oertel, M., Hempel, M., Kl{\"a}hn, T., \& Typel, S. 2017,
  \href{http://dx.doi.org/10.1103/RevModPhys.89.015007}{\JournalTitle{Rev. Mod.
  Phys.}, 89, 015007}

\bibitem[{Oppenheimer \& Volkoff(1939)}]{OpVo1939}
Oppenheimer, J.~R., \& Volkoff, G.~M. 1939,
  \href{http://dx.doi.org/10.1103/PhysRev.55.374}{\JournalTitle{Phys. Rev.},
  55, 374}

\bibitem[{{O'Shaughnessy} {et~al.}(2008){O'Shaughnessy}, {Belczynski}, \&
  {Kalogera}}]{OShaughnBelczynski08}
{O'Shaughnessy}, R., {Belczynski}, K., \& {Kalogera}, V. 2008,
  \href{http://dx.doi.org/10.1086/526334}{\JournalTitle{\apj}, 675, 566}

\bibitem[{{\"O}zel \& Freire(2016)}]{OzFr2016}
{\"O}zel, F., \& Freire, P. 2016,
  \href{http://dx.doi.org/10.1146/annurev-astro-081915-023322}{\JournalTitle{\araa},
  54, 401}

\bibitem[{Perego {et~al.}(2014)Perego, Rosswog, Cabez\'on, {et~al.}}]{PeRo2014}
Perego, A., Rosswog, S., Cabez\'on, R.~M., {et~al.} 2014,
  \href{http://dx.doi.org/10.1093/mnras/stu1352}{\JournalTitle{\mnras}, 443,
  3134}

\bibitem[{{Pian} {et~al.}(2017){Pian}, {D'Avanzo}, {et~al.}}]{PiDa2017}
{Pian}, E., {D'Avanzo}, P., {et~al.} 2017,
  \href{http://dx.doi.org/10.1038/nature24298}{\JournalTitle{Nature}, in press}

\bibitem[{Qian(2000)}]{Qian200}
Qian, Y.-Z. 2000,
  \href{http://stacks.iop.org/1538-4357/534/i=1/a=L67}{\JournalTitle{ApJL},
  534, L67}

\bibitem[{Radice {et~al.}(2016)Radice, Galeazzi, Lippuner, Roberts, Ott, \&
  Rezzolla}]{Radice:2016dwd}
Radice, D., Galeazzi, F., Lippuner, J., {et~al.} 2016,
  \href{http://dx.doi.org/10.1093/mnras/stw1227}{\JournalTitle{\mnras}, 460,
  3255}

\bibitem[{{Roberts} {et~al.}(2011){Roberts}, {Kasen}, {Lee}, \&
  {Ramirez-Ruiz}}]{RoKa2011}
{Roberts}, L.~F., {Kasen}, D., {Lee}, W.~H., \& {Ramirez-Ruiz}, E. 2011,
  \href{http://dx.doi.org/10.1088/2041-8205/736/1/L21}{\JournalTitle{ApJL},
  736, L21}

\bibitem[{{Rosswog}(2013)}]{Ro2013}
{Rosswog}, S. 2013,
  \href{http://dx.doi.org/10.1098/rsta.2012.0272}{\JournalTitle{Phil. Trans. R.
  Soc. A}, 371, 20120272}

\bibitem[{Rosswog {et~al.}(2017)Rosswog, Feindt, Korobkin, {et~al.}}]{RoFe2017}
Rosswog, S., Feindt, U., Korobkin, O., {et~al.} 2017,
  \href{http://dx.doi.org/10.1088/1361-6382/aa68a9}{\JournalTitle{Class. Quant.
  Grav.}, 34, 104001}

\bibitem[{{Rosswog} {et~al.}(1999){Rosswog}, {Liebend{\"o}rfer}, {Thielemann},
  {Davies}, {Benz}, \& {Piran}}]{RosswogLiebendorfer99}
{Rosswog}, S., {Liebend{\"o}rfer}, M., {Thielemann}, F.-K., {et~al.} 1999,
  \JournalTitle{\aap}, 341, 499

\bibitem[{{Sekiguchi} {et~al.}(2016){Sekiguchi}, {Kiuchi}, {Kyutoku},
  {Shibata}, \& {Taniguchi}}]{SeKi2016}
{Sekiguchi}, Y., {Kiuchi}, K., {Kyutoku}, K., {Shibata}, M., \& {Taniguchi}, K.
  2016,
  \href{http://dx.doi.org/10.1103/PhysRevD.93.124046}{\JournalTitle{\prd}, 93,
  124046}

\bibitem[{{Shibata} {et~al.}(2017){Shibata}, {Kiuchi}, \&
  {Sekiguchi}}]{ShKi2017}
{Shibata}, M., {Kiuchi}, K., \& {Sekiguchi}, Y.-i. 2017,
  \href{http://dx.doi.org/10.1103/PhysRevD.95.083005}{\JournalTitle{\prd}, 95,
  083005}

\bibitem[{{Siegel} \& {Metzger}(2017)}]{SiMe2017}
{Siegel}, D.~M., \& {Metzger}, B.~D. 2017, \JournalTitle{ArXiv e-prints},
  \href{http://arxiv.org/abs/1705.05473}{{\sffamily arXiv:1705.05473
  [astro-ph.HE]}}

\bibitem[{{Smartt} {et~al.}(2017)}]{Smartt2017}
{Smartt}, S.~J., {et~al.} 2017,
  \href{http://dx.doi.org/10.1038/nature24303}{\JournalTitle{Nature}, in press}

\bibitem[{{Sneden} {et~al.}(2008){Sneden}, {Cowan}, \&
  {Gallino}}]{2008ARA&A..46..241S}
{Sneden}, C., {Cowan}, J.~J., \& {Gallino}, R. 2008,
  \href{http://dx.doi.org/10.1146/annurev.astro.46.060407.145207}{\JournalTitle{\araa},
  46, 241}

\bibitem[{{Soares-Santos} {et~al.}(2017)}]{DECam}
{Soares-Santos}, M., {et~al.} 2017,
  \href{http://dx.doi.org/10.3847/2041-8213/aa9059}{\JournalTitle{ApJL}, in
  press}

\bibitem[{Tanaka \& Hotokezaka(2013)}]{TaHo2013}
Tanaka, M., \& Hotokezaka, K. 2013,
  \href{http://dx.doi.org/10.1088/0004-637X/775/2/113}{\JournalTitle{\apj},
  775, 113}

\bibitem[{Tanaka {et~al.}(2017)Tanaka, Utsumi, Mazzali, {et~al.}}]{TaUt2017}
Tanaka, M., Utsumi, Y., Mazzali, P.~A., {et~al.} 2017,
  \href{http://dx.doi.org/10.1093/pasj/psx121}{\JournalTitle{PASJ}, in press}

\bibitem[{{Tanaka} {et~al.}(2017){Tanaka}, {Kato}, {Gaigalas}, {Rynkun},
  {Radziute}, {Wanajo}, {Sekiguchi}, {Nakamura}, {Tanuma}, {Murakami}, \&
  {Sakaue}}]{Tanaka2017}
{Tanaka}, M., {Kato}, D., {Gaigalas}, G., {et~al.} 2017, \JournalTitle{ArXiv
  e-prints}, \href{http://arxiv.org/abs/1708.09101}{{\sffamily arXiv:1708.09101
  [astro-ph.HE]}}

\bibitem[{Tanvir {et~al.}(2017)}]{TanvEA2017}
Tanvir, N.~R., {et~al.} 2017,
  \href{http://dx.doi.org/10.3847/2041-8213/aa90b6}{\JournalTitle{ApJL}, in
  press}

\bibitem[{Terasawa {et~al.}(2001)Terasawa, Sumiyoshi, Kajino, Mathews, \&
  Tanihata}]{TeSu2001}
Terasawa, M., Sumiyoshi, K., Kajino, T., Mathews, G.~J., \& Tanihata, I. 2001,
  \href{http://stacks.iop.org/0004-637X/562/i=1/a=470}{\JournalTitle{\apj},
  562, 470}

\bibitem[{Troja {et~al.}(2017)Troja, Piro, van Eerten, {et~al.}}]{TrPi2017}
Troja, E., Piro, L., van Eerten, H.~J., {et~al.} 2017,
  \href{http://dx.doi.org/10.1038/nature24290}{\JournalTitle{Nature}, in press}

\bibitem[{{Valenti} {et~al.}(2017)}]{DLT40}
{Valenti}, S., {et~al.} 2017,
  \href{http://dx.doi.org/10.3847/2041-8213/aa8edf}{\JournalTitle{ApJL}, in
  press}

\bibitem[{Veitch {et~al.}(2015)Veitch, Raymond, Farr, Farr, Graff, Vitale,
  Aylott, Blackburn, Christensen, Coughlin, Del~Pozzo, Feroz, Gair, Haster,
  Kalogera, Littenberg, Mandel, O'Shaughnessy, Pitkin, Rodriguez, R\"over,
  Sidery, Smith, Van Der~Sluys, Vecchio, Vousden, \& Wade}]{PhysRevD.91.042003}
Veitch, J., Raymond, V., Farr, B., {et~al.} 2015,
  \href{http://dx.doi.org/10.1103/PhysRevD.91.042003}{\JournalTitle{Phys. Rev.
  D}, 91, 042003}

\bibitem[{Wade {et~al.}(2014)Wade, Creighton, Ochsner, Lackey, Farr,
  Littenberg, \& Raymond}]{WaCr2014}
Wade, L., Creighton, J. D.~E., Ochsner, E., {et~al.} 2014,
  \href{http://dx.doi.org/10.1103/PhysRevD.89.103012}{\JournalTitle{Phys. Rev.
  D}, 89, 103012}

\bibitem[{{Wanajo} {et~al.}(2014){Wanajo}, {Sekiguchi}, {Nishimura}, {Kiuchi},
  {Kyutoku}, \& {Shibata}}]{WanajoSekiguchi14}
{Wanajo}, S., {Sekiguchi}, Y., {Nishimura}, N., {et~al.} 2014,
  \href{http://dx.doi.org/10.1088/2041-8205/789/2/L39}{\JournalTitle{ApJL},
  789, L39}

\bibitem[{Wiringa {et~al.}(1988)Wiringa, Fiks, \& Fabrocini}]{WiFi1988}
Wiringa, R.~B., Fiks, V., \& Fabrocini, A. 1988,
  \href{http://dx.doi.org/10.1103/PhysRevC.38.1010}{\JournalTitle{Phys. Rev.
  C}, 38, 1010}

\bibitem[{{Wollaeger} {et~al.}(2017){Wollaeger}, {Korobkin}, {Fontes},
  {Rosswog}, {Even}, {Fryer}, {Sollerman}, {Hungerford}, {van Rossum}, \&
  {Wollaber}}]{WoKo2017}
{Wollaeger}, R.~T., {Korobkin}, O., {Fontes}, C.~J., {et~al.} 2017,
  \JournalTitle{ArXiv e-prints},
  \href{http://arxiv.org/abs/1705.07084}{{\sffamily arXiv:1705.07084
  [astro-ph.HE]}}

\bibitem[{Yagi \& Yunes(2017)}]{YaYu2017}
Yagi, K., \& Yunes, N. 2017,
  \href{http://dx.doi.org/10.1016/j.physrep.2017.03.002}{\JournalTitle{\physrep},
  681, 1}

\end{thebibliography}

\end{document}